%% file: TRL.tex
\documentclass[12pt,a4paper]{article}
\usepackage[utf8]{inputenc}
\usepackage{amsmath}
\usepackage{amsfonts}
\usepackage{amssymb}
\usepackage{hyphenat}
\usepackage{enumitem}
\usepackage[left=2cm,right=2cm,top=2.5cm,bottom=2.5cm]{geometry}
\usepackage{cite}
\usepackage{pdfpages}

\usepackage{slashed}
\usepackage{hyperref}
\usepackage{graphicx}
\usepackage{caption}
\usepackage{float}
\usepackage{subcaption}
\usepackage[labelformat=simple]{subcaption}

\usepackage{authblk}
\hypersetup{
	unicode=true,          
	pdftoolbar=true,        
	pdfmenubar=true,        
	pdffitwindow=false,     
	pdfstartview={FitH},    
	pdftitle={TriResonantLeptogenesis},    
	pdfauthor={},     
	pdfsubject={},   
	pdfcreator={},   
	pdfproducer={Producer}, 
	pdfkeywords={leptogenesis} {resonant leptogenesis} {Seesaw} {Baryon asymmetry} {Boltzmann equations}, 
	pdfnewwindow=true,      
	colorlinks=true,       
	linkcolor=blue,          
	citecolor=red,        
	filecolor=black,      
	urlcolor=blue           
}

\newcommand{\m}{{\bf m}}
\newcommand{\h}{{\bf h}}
\newcommand{\Hnu}{{\bf H}^\nu}

\newcommand{\Id}{{\bf 1}}

\include{macros}

 \renewcommand{\theequation}{\arabic{section}.\arabic{equation}}
\usepackage{indentfirst}

\definecolor{ao}{rgb}{0.0, 0.0, 1.0}
\newenvironment{bl}[1]{{\color{ao}  #1}}

\newcommand*{\email}[1]{%
	\footnotesize\href{mailto:#1}{\bl{#1}}
}

\title{\textbf{Tri-Resonant Leptogenesis in a Seesaw Extension of the Standard Model}}

\author[]{P. Candia da Silva\footnote{\email{pablo.candiadasilva@manchester.ac.uk}} }
\author[]{D. Karamitros\footnote{\email{dimitrios.karamitros@manchester.ac.uk}} }
\author[]{T. McKelvey\footnote{\email{thomas.mckelvey@manchester.ac.uk}} }
\author[]{A. Pilaftsis\footnote{\email{apostolos.pilaftsis@manchester.ac.uk}} }

\affil[]{
\normalsize\textit{\hspace{1.4cm}Department
 of Physics and Astronomy, University of Manchester,\newline Manchester, M13 9PL, United Kingdom}
}
 
\date{\empty}

\begin{document}

\setcounter{page}{1}

{
}	

{\let\newpage\relax\maketitle}

\maketitle

\flushbottom
\vspace{-1cm}
\begin{abstract}
  \noindent We study a class of leptogenesis models where the light
  neutrinos acquire their observed small masses by a
  symmetry-motivated construction. This class of models may naturally
  include three nearly degenerate heavy Majorana neutrinos that can
  strongly mix with one another and have mass differences comparable
  to their decay widths.  We find that such a tri-resonant heavy
  neutrino system can lead to leptonic CP asymmetries which are
  further enhanced than those obtained in the usual
  bi-resonant approximation.  Moreover, we solve the Boltzmann
  equations by paying special attention to the temperature dependence
  of the relativistic degrees of freedom of the plasma. The latter
  results in significant corrections to the evolution equations for
  the heavy neutrinos and the lepton asymmetry that have been
  previously ignored in the literature. We show the importance of
  these corrections to accurately describe the dynamical evolution of the
  baryon-to-photon ratio $\eta_B$ for heavy neutrino masses at and
  below $100~\GeV$, and demonstrate that successful lepto\-genesis at
  lower masses can be significantly affected by the variation of the
  relativistic degrees of freedom.  The parameter space for the
  leptogenesis model is discussed, and it could be probed in future
  experimental facilities searching for charged lepton flavour
  violation and heavy neutrinos in future $Z$-boson factories.
\end{abstract}

\newpage
\tableofcontents

\section{Introduction}\label{sec:Intro}
\setcounter{equation}{0}

Observations done by the Wilkinson Microwave Anisotropy Probe (WMAP)
and the Planck observatory indicate that the extent of the Baryon
Asymmetry of the Universe (BAU) amounts
to~\cite{Planck:2018vyg,Fields:2019pfx} 
\begin{equation}
\eta_B^{\rm CMB} = 6.104\pm 0.058 \times 10^{-10}.
\end{equation}
Hence, explaining the observed BAU has been one of the central themes
of Particle Cosmology for decades. The existence of this non-zero BAU
is one of the greatest pieces of evidence for physics beyond the
Standard Model (SM).  In the SM, the neutrinos are strictly massless,
and so this runs contrary to the observations of neutrino
oscillations\cite{Ahmad:2001an,Ahmad:2002jz,Fukuda:1998mi}, which only
exist for massive neutrinos. A minimal resolution to this problem will be
to include additional heavy neutrinos which are singlets under the SM
gauge group:
${\textrm{SU}(3)_c\times \textrm{SU}(2)_L \times \textrm{U}(1)_Y}$.
These additional neutrinos are permitted to have large masses due to
the inclusion of a Majorana mass term which violates lepton number,
$L$, by two units. They also provide a mechanism to render the SM
neutrinos massive, whilst ensuring that the generated mass is small in
scale through the famous seesaw mechanism~\cite{Minkowski:1977sc,
  GellMann:1980vs,Yanagida:1979as,Mohapatra:1979ia}.  On the other
hand, the spacetime expansion of the FRW Universe provides a
macroscopic arrow of cosmic time $t$, as well as Sakharov's necessary
out-of-thermal equilibrium condition~\cite{Sakharov:1967dj} needed to
potentially generate a large lepton-number asymmetry. This asymmetry
is then rapidly converted into a baryon asymmetry through
$(B+L)$-violating sphaleron transitions while the temperature of the
Universe remains above the temperature
$\Tsph \approx 132~\GeV$, after which these sphaleron transitions
become exponentially suppressed. This mechanism is commonly referred
to as \textit{leptogenesis}~\cite{FukYan:1986}.

A particularly interesting framework of leptogenesis is
\textit{Resonant Leptogenesis}
(RL)\cite{Pilaftsis:1997jf,Pilaftsis:2003gt}, which permits Majorana
mass scales far lower than those that occur in typical Grand Unified
Theory (\textrm{GUT}) models of leptogenesis~\cite{FukYan:1986,DiBari:2020plh}. In
RL models, the CP violation generated is greatly enhanced through
the mixing of nearly degenerate heavy Majorana neutrinos $N_\alpha$, provided
\begin{equation*}
    \left| m_{N_\alpha} - m_{N_\beta} \right| \sim \frac{\Gamma_{N_{\alpha,\beta}}}{2}\;,
  \end{equation*}
  where $m_{N_\alpha}$ and $\Gamma_{N_\alpha}$ are the masses and the
  decay widths of  $N_\alpha$, respectively.  This
  mass arrangement in turn permits the generation of appreciable BAU
  at sub-$\TeV$ masses~\cite{Pilaftsis:1997jf,Pilaftsis:1998pd}, in
  agreement with neutrino oscillation
  parameters~\cite{Pilaftsis:2003gt,Pilaftsis:2005rv}.

In this paper we study a class of leptogenesis models that may
naturally include three nearly degenerate heavy Majorana neutrinos
which can strongly mix with one another and have mass differences
comparable to their widths. We compute the leptonic CP asymmetries
generated in such a tri-resonant heavy neutrino system, to find that
their size is further enhanced in comparison to those that were
naively determined in the usually considered bi-resonant
approximation. Accordingly, this enhanced mechanism of leptogenesis
will be called Tri-Resonant Lepto\-genesis~(TRL). In the context of
models realising TRL, our aim is to find neutrino Yukawa couplings
whose size lies much higher than the one expected from a typical
seesaw scenario, whilst still achieving the observed BAU. To this end,
we solve the Boltzmann equations (BEs) that describe the evolution of
heavy neutrino and lepton-asymmetry number densities before\- the
sphaleron freeze-out temperature, after including decay and scattering
collision terms.  An important\- novelty of the present study is to assess the
significance of the temperature dependence of the relativistic degrees
of freedom (dofs) in the plasma. Finally, we analyse observables
of charged Lepton Flavour Violation (cLFV) that could be tested in
current and projected experiments${}$, such as ${\mu \rightarrow eee}$ at
\textrm{Mu3e}~\cite{Blondel:2013ia}, ${\mu \rightarrow e \gamma}$ at
\textrm{MEG} ~\cite{MEG:2016leq,MEGII:2018kmf}, coherent
$\mu\rightarrow e$ conversion at COMET~\cite{Moritsu:2022lem} and
PRISM~\cite{BARLOW201144}, as well as matching the observed light
neutrino mass
constraints~\cite{Ahmad:2001an,Ahmad:2002jz,Fukuda:1998mi}.

In our analysis we will not specify the origin of the structure of the
Majorana-mass and the neutrino Yukawa matrices.  But we envisage a
high-scale SO(3)-symmetric mass spectrum for the heavy Majorana
neutrinos, possibly of the order of GUT
scale~\cite{Deppisch:2010fr,Pilaftsis:2015bja}, which is broken by
renormalisation-group (RG) and new-physics threshold effects.
Following a less constrained approach to model-building, we also
assume an approximate $\mathbb{Z}_6$-symmetric texture for the entries
of the neutrino Yukawa matrix.  Such a construction enables the
generation of the observed small neutrino masses, without imposing the
expected seesaw suppression on the neutrino Yukawa parameters for
heavy neutrino masses at the electroweak scale.

The layout of the paper is as follows. In Section~\ref{sec:Z_6_seesaw}
we describe the minimal extension of the SM that we will be studying,
and introduce the flavour structure of its leptonic Yukawa sector. In
Section~\ref{sec:LowEnObs} we specify the light neutrino mass spectrum
for our analysis, and present the cLFV observables one may expect to
probe in this model, such as ${\mu \rightarrow e\gamma}$,
${\mu \rightarrow eee}$, and coherent $\mu\to e$ conversion in
nuclei. In Section~\ref{sec:Leptogen} we explore the different aspects
of leptogenesis, notably the CP violation generated in RL and TRL
scenarios and derive the relevant set of BEs, upon which our numerical
estimates are based. This set of BEs is solved including contributions
from chemical potentials while crucially preserving the temperature
dependence of the key parameter, denoted later as~$h_{\rm eff}(T)$,
that describes the variation of the relativistic dofs with~$T$. In
Section~\ref{sec:approx} we present approximate solutions to the BEs,
which will help us to shed light on the attractor properties of our
fully-fledged numerical estimates. In Section~\ref{sec:results} we
give a summary of our numerical results, including evolution plots for
the BAU and comparisons with observable quantities. Finally,
Section~\ref{sec:conclusions} summarises our conclusions and discusses
possible future directions. Some technical aspects of our study have been
relegated to Appendices~\ref{app:BE_dh}, \ref{app:formfactors}
and~\ref{app:deltah_values}.

\vfill\eject

\section{Seesaw Extension of the Standard Model}\label{sec:Z_6_seesaw}
\setcounter{equation}{0}

We adopt the framework of the
conventional seesaw extension of the SM. This extension requires the
addition of $n\geq2$ right-handed neutrinos, which are singlets under the SM
gauge group, and have lepton number $L_{\nu_R}=1$. Given this particle
content and quantum number assignments, the Lagrangian of the right-handed neutrino sector reads: 
\begin{align}
    \mathcal{L}_{\nu_R} =i\overline{\nu}_R\slashed{\partial}\nu_R -\left( \overline{L}\,\h^\nu\tilde{\Phi}\,\nu_R + \frac{1}{2}\overline{\nu}^C_R\,\m_M\nu_R + {\rm H.c.\, }\right)\label{eq:seesaw_lagrangian}.
\end{align}
Here, $L_i=(\nu_{iL},e_{iL})^{\sf T}$, with $i=1,2,3$, denote the left-handed lepton doublets, while $\nu_{\alpha R}$, with $\alpha=1,...,n$, are the right-handed neutrino fields. The matrices $\h^\nu$ and $\m_M$ are the neutrino Yukawa and the Majorana mass matrices, respectively, and $\tilde{\Phi}$ is the weak isospin conjugate of the Higgs doublet $\Phi$. Note that we reserve bold face for matrices in flavour space, and assume the implicit contraction of flavour space indices.

Without loss of generality, we assume that the Majorana mass matrix is diagonal, in which case we may recast the Lagrangian~\refs{eq:seesaw_lagrangian} in the unbroken phase as 
\begin{align}
\mathcal{L}_{\nu_R} =i\overline{N}\slashed{\partial}N-\left( \overline{L}\,\h^\nu\tilde{\Phi}\,P_R N + {\rm H.c.}\right) - \frac{1}{2}\overline{N}\,\m_MN,\label{eq:seesaw_lagrangian_mass}
\end{align}
where $P_{R/L} = \frac{1}{2}\left( \Id_4 \pm \gamma_5 \right)$ is the
right-/left- chiral projector, $\Id_n$ is the $n \times n$ identity
matrix, $\m_M={\rm diag}(m_{N_1},...,m_{N_n})$, and
$N_\alpha = \nu_{\alpha R} + \nu^C_{\alpha R}$ are the mass-eigenstate
Majorana spinors associated to the right-handed neutrinos.

In the broken phase, this picture changes by the mixing between singlet and left-handed neutrinos. The mass eigenstates are particular combinations of the weak eigenstate neutrinos, given by 
\begin{align}
P_R \begin{pmatrix}
\nu \\
N
\end{pmatrix}=
\begin{pmatrix}
U_{\nu\nu_L^C} & U_{\nu \nu_R}\\
U_{N\nu_L^C} & U_{N \nu_R}
\end{pmatrix}
\begin{pmatrix}
\nu_L^C\\
\nu_R
\end{pmatrix}\; ,
\end{align}
where $\nu_{1,2,3}$ are the light neutrino mass eigenstates and $U$ is
a $(3+n)\times(3+n)$ unitary matrix that diagonalises the neutrino
mass matrix (see Section \ref{sec:z6}). The subscripts, $\nu^C_L$ and
$\nu_R$, on its sub-blocks indicate the possible components of the
right-handed chirality projection of each mass eigenstate, represented
here as vector columns $\nu$ and $N$. Following the notation
of~\cite{Pilaftsis:1991ug}, we may then write the Lagrangian for the
charged current interaction of the heavy neutrinos as 
\begin{align}
\mathcal{L}^W_{\rm int} = -\frac{g_w}{\sqrt{2}}W^-_\mu
  \overline{e}_{iL} B_{i\alpha}\gamma^\mu P_L N_\alpha + {\rm H.c.}\,, 
\end{align}
where $g_w$ is the gauge coupling associated to the SU(2)$_L$ group, and
\begin{align}
B_{i\alpha}\simeq \xi_{i\alpha}= (\m_D\m^{-1}_M)_{i\alpha}\label{eq:light-heavy_mixing}
\end{align}
is the light-to-heavy neutrino mixing at first order in the expansion
of the matrix-valued para\-meter~$\xi$~\cite{Pilaftsis:1991ug}. In the
following, we assume that the charged lepton Yukawa matrix is diagonal, and
hence $B_{i\alpha}=(U_{\nu \nu_{R}})_{i\alpha}$. 
At first order in $\xi$, the effective light neutrino mass matrix,
$\m^\nu$, follows the well-known seesaw
relation~\cite{GellMann:1980vs} 
\begin{align}
\m^\nu = -\m_D \m^{-1}_M \m^{\sf T}_D \; ,\label{eq:tree_level_mass}
\end{align}
where $\m_D=\h^\nu v/\sqrt{2}$ is the Dirac mass matrix, and $v\simeq 246~\GeV$ is the vacuum expectation value (VEV) of the Higgs field. By virtue of this relation, it is apparent that a Dirac mass matrix at a scale
\begin{equation}
    |\!|\m_D|\!| \equiv \sqrt{\textrm{Tr}\left[ \mathbf{m}_D^\dagger \mathbf{m}_D \right]} \approx v \;,
\end{equation}
would in principle require GUT scale heavy neutrinos, which means that
any impact of the singlet neutrino sector on experimental signatures
would be beyond the realm of observation. This  motivates the search
for new model building strategies to explain sub-${\rm eV}$ light
neutrinos, whilst maintaining agreement with light neutrino data and
other low energy experiments. 

\subsection{Neutrino Flavour Model}
\label{sec:z6}

In order to explain the smallness of neutrino masses, we investigate
scenarios where the neutrino mass matrix is naturally small,
preferably arising from the subtle breaking of a symmetry. When this
symmetry is exact,~\refs{eq:tree_level_mass} vanishes identically,
given by the $3\times 3$ null matrix, $\mathbf{0}_3$, \ie 
\begin{align}
    \m_D\, \m^{-1}_M \,\m^{\sf T}_D\, =\, {\bf0}_3 \;.
\end{align}
If we consider a singlet neutrino sector with a nearly degenerate mass
spectrum, this is approximately equivalent to require that prior to
the breaking of the symmetry the leading Yukawa matrix, $\h^\nu_0$,
satisfies the condition 
\begin{align}
    \h^\nu_0 \, \h^{\nu\sf T}_0\, =\, {\bf0}_3\;.
    \label{eq:zeromassyuk}
\end{align}
Considering a model with three right-handed neutrinos, this motivates
the following structure for the leading neutrino Yukawa matrix: 
\begin{align}
    \h^\nu_0=\begin{pmatrix}
        a & a\,\omega & a\,\omega^2\\
        b & b\,\omega & b\,\omega^2\\
        c & c\,\omega & c\,\omega^2
    \end{pmatrix}\;,
    \label{eq:z6yukawa}
\end{align}
where the parameters $a$, $b$, and $c$ are in general real, and $\omega$ is the
generator of the $\mathbb{Z}_6$ group, $\omega = \exp(\pi i/3)$. We
remark that this choice is not unique, and the vanishing of the light
neutrino mass matrix may be realised through other
constructions of the neutrino Yukawa matrix. For example, one could
replace the $\mathbb{Z}_6$ element $\omega$ with the $\mathbb{Z}_3$
element $\omega^\prime=\exp(-2\pi i/ 3)$. However, for concreteness, we
select the $\mathbb{Z}_6$-symmetry  realisation for our analysis. 

Evidently, the flavour structure of~\refs{eq:z6yukawa} has to be
perturbed in order to reproduce the observed neutrino oscillation
phenomenon, which requires massive neutrinos. Even though $\h^\nu_0$
as given by~\refs{eq:z6yukawa} is rank one, a perturbation $\delta
\h^\nu$ such that rank($\delta\h^\nu)\geq2$ is sufficient to explain
neutrino oscillations, as long as the following condition is enforced: 
\begin{align}
    (\h^\nu_0 + \delta\h^\nu)\,\m_M^{-1}\,(\h^\nu_0 + \delta\h^\nu)^{\sf
  T}\, =\, \frac{2}{v^2}\,\m^\nu\;. 
    \label{eq:numassconstraint}
\end{align} 
where $\m^\nu$ is a $3\times 3$ complex and symmetric matrix.  Taking
$a$, $b$, $c$, and the singlet neutrino spectrum as input
parameters,~\refs{eq:numassconstraint} defines a set of $12$
constraints for the entries of the perturbation
matrix~$\delta\h^\nu$. The solutions to~\refs{eq:numassconstraint}
have to satisfy a further condition, which is
$|\delta\h^\nu_{ij}|/|(\h^\nu_0)_{kl}|\ll1$, with $i,j,k,l=1,2,3$.
More generally, the zero mass condition of~\refs{eq:zeromassyuk} can
be enforced when the Majorana mass matrix $\m_M$ is not proportional
to the identity, or even in the case when loop corrections to the
tree-level seesaw relation of~\refs{eq:tree_level_mass} are
considered. This gives us complete control over the loop corrections
to the light-neutrino mass matrix at all orders. For example, we can
incorporate one-loop corrections to $\mathbf{m}^\nu$
\cite{Pilaftsis:1991ug} by modifying the tree-level zero mass
condition as follows~\cite{Dev:2012sg}:
\begin{align}
    \h^\nu_0\left[\m^{-1}_M - \frac{\alpha_w}{16\pi
  M^2_W}\m^{\dagger}_Mf(\m_M\m^\dagger_M)\right]\h^{\nu\sf
  T}_0=\,\mathbf{0}_3\;, 
    \label{eq:one_loop_zero_mass}
\end{align}
where
\begin{align}
    f(\m_M\m^\dagger_M)=\frac{M^2_H}{\m_M\m^\dagger_M -
  M^2_H\Id_3}\ln\left(\frac{\m_M\m^\dagger_M}{M^2_H}\right) +
  \frac{3M^2_Z}{\m_M\m^\dagger_M -
  M^2_Z\Id_3}\ln\left(\frac{\m_M\m^\dagger_M}{M^2_Z}\right)\;. 
    \label{eq:loop_factor_f_def}
\end{align}
In the above, $\alpha_w\equiv g_w^2/(4\pi)^2$ is the
electroweak-coupling parameter, and $M_W$, $M_Z$, and $M_H$ are the
masses of the $W$, $Z$, and Higgs bosons, respectively.  Redefining
the quantity inside the square brackets
in~\refs{eq:one_loop_zero_mass} as an effective inverse Majorana mass,
$\overline{\m}_M^{-1}$, the restrictions can be recast as
\begin{align}
    \h^\nu_0\overline{\m}_M^{-1}\h^{\nu\sf T}_0=\,\mathbf{0}_3\;.
    \label{eq:recast_zero-mass_constraint}
\end{align}
This can be further simplified by rescaling the columns of the Yukawa
matrix using the definition
\begin{align}
    \Hnu_0 =\h^\nu_0\,\overline{\m}^{-1/2}_M\;,
    \label{eq:Htoh}
\end{align}
which leads to
\begin{align}
    \Hnu_0 \, {\Hnu_0}^{\sf T}=\,\mathbf{0}_3\;.
    \label{eq:H0_zero-mass_constraint}
\end{align}
This results in the same condition of~\refs{eq:zeromassyuk} but this
time for a rescaled Yukawa matrix, $\Hnu_0$, with dimensions of
(mass)$^{-1/2}$. This shows that even for appreciable mass splittings
between the singlet neutrinos and with the inclusion of loop
corrections to the neutrino mass matrix, the Yukawa matrix can always
be chosen in such a way that the neutrinos are massless by taking 
\begin{align}
    \Hnu_0 = \begin{pmatrix}
        \ell_1 & \ell_1\,\omega & \ell_1\,\omega^2\\
        \ell_2 & \ell_2\,\omega & \ell_2\,\omega^2\\
        \ell_3 & \ell_3\,\omega & \ell_3\,\omega^2
    \end{pmatrix}\;,
    \label{eq:rescaledyukawa}
\end{align}
where $\ell_{1,2,3}$ are real parameters. The dimensionless Yukawa
matrix, $\h^\nu_0$, can be found using~\refs{eq:Htoh}, and as
explained previously, its structure can then be perturbed to reproduce
the observed neutrino mass matrix $\m^\nu$. Here we will not address
the origin of the texture of the neutrino Yukawa matrix $\h^\nu_0$,
but it can be the subject of future studies on model building. 

It is worthy to mention that the neutrino mass matrix is model
dependent, and its relation to the observable parameters measured in
neutrino oscillation experiments is given by 
\begin{align}
    \m^\nu = U_\textrm{PMNS}^{\sf T}\,\widehat{\m}^\nu U_\textrm{PMNS}\;,
    \label{eq:mnudiag}
\end{align}
where $U_\textrm{PMNS}$ is the PMNS lepton mixing
matrix~\cite{Pontecorvo:1957qd,Maki:1962mu} and $\widehat{\m}^\nu={\rm
  diag}(m_1,m_2,m_3)$, in which $m_{1,2,3}$ are the light neutrino
masses. The matrix $U_\textrm{PMNS}$ performs the Takagi factorisation
\cite{auto:15,takagi:25} when applied to the light neutrino mass
matrix. If the Yukawa matrix of the charged leptons is assumed to be
diagonal, $U_\textrm{PMNS}$ parameterises the flavour mixing in
charged current interactions of the leptonic sector. The experimental
values of the parameters involved in~\refs{eq:mnudiag} are discussed
in the next section. 

\section{Low Energy Observables}\label{sec:LowEnObs}
\setcounter{equation}{0}
The observation of flavour neutrino oscillations at
Super-Kamiokande~\cite{Fukuda:1998mi} and the Sudbury Neutrino
Observatory~\cite{Ahmad:2001an,Ahmad:2002jz} provides definite
evidence of their massive nature. The resulting neutrino oscillation
parameters offer strong constraints on the neutrino model parameters,
which we discuss in this section. In addition, we present the formulae
for the rates of selected charged cLFV processes, namely
$\mu\rightarrow e\gamma$, $\mu\rightarrow eee$ and coherent
$\mu\rightarrow e$ conversion in nuclei, which can be distinctive
signatures of Majorana neutrino models, and are crucially dependent on
the light-to-heavy neutrino mixing parameter presented
in~\refs{eq:light-heavy_mixing}.

\subsection{Neutrino Oscillation Data}
In order to incorporate the neutrino mass constraints into our model,
we follow the procedure outlined in Section~\ref{sec:z6}. We neglect
the non-unitarity effects that arise due to light-to-heavy neutrino
mixing, and without loss of generality, we assume that the charged
lepton Yukawa matrix, $\h^\ell$, is diagonal. With the first
assumption in mind, the matrix $U_\textrm{PMNS}$ can be parameterised
as follows \cite{BILENKY1980495, Schechter:1980gr}:
\begin{equation}
    U_\textrm{PMNS}\ =\ \begin{pmatrix}
        c_{12}c_{13} & s_{12}c_{13} & s_{13}e^{-i\delta} \\
        -s_{12}c_{23}-c_{12}s_{23}s_{13}e^{i\delta} & c_{12}c_{23}-s_{12}s_{23}s_{13}e^{i\delta} & s_{23}c_{13}\\
        s_{12}c_{23}-c_{12}c_{23}s_{13}e^{i\delta} & -c_{12}s_{23}-s_{12}c_{23}s_{13}e^{i\delta} & c_{23}c_{13}
    \end{pmatrix}\times \text{diag}(e^{i\alpha_1/2},e^{i\alpha_2/2},1)\;,
    \label{eq:U_def}
\end{equation}
where $c_{ij} = \cos\theta_{ij}$ and $s_{ij} = \sin\theta_{ij}$ are the cosines and sines of
the neutrino mixing angles, $\delta$ is the so-called Dirac  phase, and
$\alpha_{1,2}$ are the Majorana phases. Together with the neutrino
squared mass differences $\Delta m^2_{21}\equiv m^2_2-m^2_1$ and
$\Delta m^2_{31}\equiv m^2_3-m^2_1$, these angles and the Dirac phase
comprise the light neutrino oscillation data. 

The values of these parameters are experimentally bounded with the
exception of the absolute neutrino mass scale, characterised by
min($m_{1,3}$), and the sign of $\Delta m^2_{31}\equiv m^2_3 - m^2_1$,
which requires the distinction between the normal
($\Delta m^2_{31}>0$) and inverted ($\Delta m^2_{31}<0$) ordering
hypotheses. For our numerical estimates, we use the latest best fit
values for the neutrino oscillation parameters~\cite{deSalas:2020pgw}:
\begin{align}
    &\Delta m^2_{21}\equiv m^2_2 - m^2_1 = 7.50\times
    10^{-5}\,(\text{eV})^2,\quad\Delta m^2_{31}\equiv m^2_3 - m^2_1
    = 2.55\times 10^{-3} \,(\text{eV})^2,\\[2mm] 
    &\theta_{12} = 34.3^{\circ},\quad \theta_{23} =
    49.26^{\circ} ,\quad\theta_{13} = 8.58^{\circ},\quad
    \delta = 194^{\circ}.
    \label{eq:neutrino_params} 
\end{align}
Since the experimental data allows a massless neutrino, for definiteness we work under the hypothesis that $m_1 = 0$ and the light neutrino spectrum follows normal ordering. Likewise, for the unconstrained Majorana phases, we set  $\alpha_{1,2} = 0$. For relevant tri-resonant benchmarks, we provide the relevant $\delta \h^{\nu}$ values, which reproduce the light neutrino data in Appendix~\ref{app:deltah_values}.   

\subsection{Lepton Flavour Violation}
In the seesaw extension of the SM, the leading order contributions to
cLFV processes appear at the one-loop level \cite{Cheng:1980tp}. For
the radiative decays of our interest, the expressions for the
branching ratios are given by~\cite{Ilakovac:1994kj} 
\begin{align}
  {\rm BR}(\mu\to e\gamma)  &= \frac{\alpha_w^3 s_w^2}
  {256\pi^2} \frac{m_\mu^4}{M_W^4}\: \frac{m_\mu}{\Gamma_\mu}
  \left|G_\gamma^{\mu e}\right|^2, \label{eq:brmue}\\
  {\rm BR}(\mu\to eee)  &= \frac{\alpha_w^4}{24576 \pi^3}\:
  \frac{m_\mu^4}{M_W^4}  \frac{m_\mu}{\Gamma_\mu}
  \left\{2 \left|\frac{1}{2}F_{\rm Box}^{\mu eee}
      +F_Z^{\mu e}-2s_w^2\left(F_Z^{\mu e}-F_\gamma^{\mu e}\right)
    \right|^2 \right. \nonumber \\
  & \left. + \ 4s_w^4\left|F_Z^{\mu e}-F_\gamma^{\mu e}\right|^2 
     +  16s_w^2 \,\Re e\left[\left(F_Z^{\mu e}
        +\frac{1}{2}F_{\rm Box}^{\mu eee}\right)
      G_\gamma^{\mu e*}\right] \right. \label{eq:brmu3e}\\
  \ & \left.
    - \ 48s_w^4\, \Re e\,\left[\left(F_Z^{\mu e}
        -F_\gamma^{\mu e}\right)G_\gamma^{\mu e*}\right] 
     +  32 s_w^4\left|G_\gamma^{\mu e}\right|^2
    \left[\ln\left(\frac{m_\mu^2}{m^2_e}\right)-\frac{11}{4}\right]
  \right\} \;, \nonumber
\end{align} 
where $s_w\equiv\sin\theta_w$ is the sine of the weak angle, $m_e$ is
the mass of the electron, and $m_\mu$ and $\Gamma_\mu$ are the muon
mass and width. The form factors are  defined in
Appendix~\ref{app:formfactors}. It is worth mentioning that other cLFV
decays involving $\tau$ leptons are also allowed, but we ignore them
in the discussion of our results since the experimental bounds that
apply to those processes are far weaker in the parameter space of 
interest to us. 

The rate for the $\mu\to e$ conversion in an atomic nucleus $_Z^A X$
is given by~\cite{Alonso:2012ji} 
\begin{eqnarray}
  R^X_{\mu\to e} \ = \ \frac{2G_F^2\alpha_w^2m_\mu^5}
  {16\pi^2\Gamma_{\rm capt}} \ \left|4V^{(p)} 
    \left(2\tilde{F}_u^{\mu e} + \tilde{F}_d^{\mu e}\right) 
    + 4V^{(n)}\left(\tilde{F}_u^{\mu e} 
      + 2\tilde{F}_d^{\mu e}\right)
    + \frac{s_w^2}{2e}G_\gamma^{\mu e}D\right|^2 \; , 
  \label{eq:ratemue}
\end{eqnarray}
where $G_F$ is Fermi's constant, $e=g_w s_w$ is charge of the electron, 
$\Gamma_{\rm capt}$ is the nuclear capture rate, and $V^{(p)},V^{(n)},D$ 
are numerical estimations of the overlap integrals involved in the calculation of the conversion rate~\cite{Kitano:2002mt}. For the nuclei of our interest, Table~\ref{tab:nuclear} presents the numerical values of these parameters.  
The form factors
$\tilde{F}_q^{\mu e}$ $(q=u,d)$ in~\refs{eq:ratemue} are defined as
\begin{eqnarray}
  \tilde{F}_q^{\mu e} \ = \ Q_qs_w^2F_\gamma^{\mu e} \;
  + \; \left(\frac{I^q_{3}}{2}-Q_qs_w^2\right)F_Z^{\mu e} \;
  + \; \frac{1}{4}F_{\rm Box}^{\mu e qq} \; ,
\end{eqnarray}
where $Q_u=2/3$, $Q_d=-1/3$ refer to the electric charges of up- and down-type quarks, and $I^u_{3}=1/2$, $I^d_{3}=-1/2$ denote the third component  of their weak isospin. The corresponding form factors can be found in Appendix~\ref{app:formfactors}.
\begin{table}[t!]
  \begin{center}
    \begin{tabular}{ccccc}\hline\hline
      Nucleus ($_Z^A X$) & $V^{(p)}$ & $V^{(n)}$ & $D$ &
      $\Gamma_{\rm capt}~(10^6~{\rm s}^{-1})$\\ \hline
      $_{13}^{27}$Al & 0.0161 & 0.0173 & 0.0362 & 13.45\\
      $_{22}^{48}$Ti & 0.0396 & 0.0468 & 0.0864 & 2.59\\
      $_{79}^{197}$Au & 0.0974 & 0.146 & 0.189 & 13.07\\
      \hline\hline
    \end{tabular}
  \end{center}
\caption{Overlap integrals and muon capture rates for the nuclei of the elements used in the relevant experiments.}\label{tab:nuclear}
\end{table}

The search for cLFV is a prominent experimental endeavour, and there are several facilities that operate with the aim of finding a conclusive hint for this class of transitions. Despite the non-observation of these signals, experimental efforts have lead to stringent bounds on the parameter space of Majorana neutrino models, which are reflected by the current upper limits
\begin{align}
    {\rm BR}(\mu\rightarrow e\gamma)&< 4.2\times10^{-13}\qquad \text{MEG~\cite{MEG:2016leq}}\;,\nn\\
    {\rm BR}(\mu\rightarrow eee)&< 1.0\times10^{-12}\qquad \text{SINDRUM~\cite{SINDRUM:1987nra}}\;,\label{eq:current_constraints}\\
R^{\rm Au}_{\mu\rightarrow e}&< 7.0\times10^{-13}\qquad \text{SINDRUM~\cite{SINDRUMII:2006dvw}}\;.\nn
\end{align}
These limits are expected to be improved by a few orders of magnitude
in the near future. There is a new generation of experiments that are
either starting to take data, under construction, or in the
proposal/design stage. Among them, we should mention MEG-II, COMET, Mu3e,
Mu2e and PRISM, with the following projected sensitivities: 
\begin{align}
    {\rm BR}(\mu\rightarrow e\gamma)&< 6\times10^{-14}\qquad \text{MEG
                                      II~\cite{MEGII:2018kmf}},\nn\\ 
    {\rm BR}(\mu\rightarrow eee)&< 10^{-16}\qquad \text{\quad\,\,
                                  Mu3e~\cite{Blondel:2013ia}},\nn\\ 
    R^{\rm Al}_{\mu\rightarrow e}&< 3\times10^{-17}\qquad
                                   \text{Mu2e~\cite{DiFalco:2022kqz}},\label{eq:projected_sensitivities}\\ 
    R^{\rm Al}_{\mu\rightarrow e}&< 10^{-17}\;\,\,\quad\qquad
                                   \text{COMET~\cite{Moritsu:2022lem}},\nn\\ 
    R^{\rm Ti}_{\mu\rightarrow e}&< 10^{-18}\qquad \text{\quad\,\,
                                   PRISM~\cite{BARLOW201144}}.\nn 
\end{align}
These projections will be compared with the cLFV rates as predicted by
our leptogenesis model to assess its testability in the foreseeable
future. 

\subsubsection{Non-Zero Leptonic CP Phases in cLFV Processes}
\label{sec:lCPV}
Here we examine the impact of leptonic CP phases on cLFV processes for
our class of seesaw models. It was argued in~\cite{Abada:2021zcm} that
the existence of non-zero leptonic CP phases may have a substantive
impact on the rate of cLFV processes through the interference terms
involving the mixing $B_{\ell \alpha}$. Following a procedure similar
to~\cite{Abada:2021zcm}, we write the elements of $B_{i \alpha}$ as a
magnitude $s_{i \alpha}$ and a phase $\varepsilon_{i \alpha}$. Thus,
the terms that appear in the observable quantities are 
\begin{equation}
    \sum_{\alpha=1}^3 B_{i\alpha} B_{j\alpha}^* = \sum_{\alpha = 1}^3 s_{i \alpha} s_{j\alpha}e^{i(\varepsilon_{i\alpha} - \varepsilon_{j\alpha})}= \sum_{\alpha = 1}^3 s_{i \alpha} s_{j\alpha}e^{i\Delta^{ij}_\alpha}\;,
    \label{eq:B_obs_def}
\end{equation}
where we have introduced the CP phases $\Delta^{ij}_{\alpha} = \varepsilon_{i\alpha} - \varepsilon_{j\alpha}$. These CP phases are expected to be small and can easily be extracted by taking the ratio of imaginary to real parts of the mixing, \ie 
\begin{equation}
    \frac{\Im m \left\{ B_{i\alpha} B_{j\alpha}^* \right\}}{\Re e \left\{B_{i\alpha} B_{j\alpha}^* \right\}} = \tan \left(\Delta_\alpha^{ij} \right) \approx \Delta_\alpha^{ij} \; .
    \label{eq:im_re_B_ratio}
\end{equation}
For the model introduced in Section~\ref{sec:Z_6_seesaw}, the heavy neutrino masses are nearly degenerate and the elements of the mixing matrix are all of similar scale. Therefore, the observable quantities may be approximated by taking the masses to be exactly degenerate and letting ${s_{i \alpha} \approx s_{i1}}$ for all $\alpha$. Under these simplifications, the variations in the cLFV observables are captured in the value of
\begin{equation}
    \left| \sum_{\alpha=1}^3 B_{i\alpha}B_{j\alpha}^* \right|^2 \approx s_{i1}^2 s_{j1}^2 \sum_{\alpha,\beta =1}^3 \cos \left( \Delta_\alpha^{ij}  - \Delta_\beta^{ij}  \right) \;.
    \label{eq:sub_abs_B_approx}
\end{equation}
Then, the observed deviation due to the existence of non-zero leptonic CP phases may be written as
\begin{equation}
    D^{ij} = 1 - \frac{\left| \sum_{\alpha=1}^3 B_{i\alpha}B_{j\alpha}^* \right|^2}{\left| \sum_{\alpha=1}^3 B_{i\alpha}B_{j\alpha}^* \right|^2_{\Delta_i = 0}} \approx 1 - \frac{1}{9} \sum_{\alpha,\beta =1}^3  \cos \left( \Delta_\alpha^{ij}  - \Delta_\beta^{ij}  \right) \;.
    \label{eq:D_ij_def}
\end{equation}
In the case of small leptonic CP phases, it can be seen that the deviation in the rate of cLFV processes away from the CP conserving rate may be given by
\begin{equation}
    D^{ij} \approx \frac{1}{9}\left[\left( \Delta_1^{ij}  - \Delta_2^{ij}  \right)^2 + \left( \Delta_1^{ij}  - \Delta_3^{ij}  \right)^2 + \left( \Delta_2^{ij}  - \Delta_3^{ij}  \right)^2\right],
    \label{eq:D_ij_small_CP_approx}
\end{equation}
and so the observed deviation is itself a small effect.

In the context of the $\mathbb{Z}_6$ motivated model we have
presented, the quantity $B_{i \alpha}B_{j\alpha}^*$ is completely real
at lowest order, and therefore the leptonic CP phases are identically
zero. Therefore, in order to have non-zero leptonic CP phases, we need
to include the symmetry breaking term~$\delta \mathbf{h}^\nu$. It is
then not easy to verify that up to leading order in the perturbations,
$\delta \mathbf{h}^\nu$, the relevant leptonic CP phases are given by
\begin{equation}
    \Delta^{ij}_\alpha \approx \frac{\Im m \left\{(\mathbf{h}^\nu_0
        \mathbf{m}_M^{-1})_{i
          \alpha}(\delta\mathbf{h}^\nu\mathbf{m}_M^{-1})_{j \alpha}^*
        +(\delta\mathbf{h}^\nu\mathbf{m}_M^{-1})_{i \alpha}
        (\mathbf{h}^\nu_0 \mathbf{m}_M^{-1})_{j
          \alpha}^*\right\}}{(\mathbf{h}^\nu_0 \mathbf{m}_M^{-1})_{i
        \alpha}(\mathbf{h}^\nu_0 \mathbf{m}_M^{-1})_{j\alpha}^*}\;. 
    \label{eq:Delta_ij_Z6_approx}
\end{equation}
Hence, $|\Delta_\alpha^{ij}| \sim |\!|\delta
\mathbf{h}^\nu|\!|/|\!|\mathbf{h}^\nu_0|\!| \ll 1$. We may therefore
expect the deviation away from the CP conserving cLFV observables to
be very small in magnitude, $D^{ij} \sim
|\!|\delta\mathbf{h}^\nu|\!|^2/(9|\!|\mathbf{h}^\nu_0|\!|^2)$ . For
the generic scenarios listed in Appendix \ref{app:deltah_values}, one
finds a deviation of $D^{ij} \sim 10^{-3}$, so any CP effect will be
difficult to observe for the TRL models under study.

\section{Tri-Resonant Leptogenesis}\label{sec:Leptogen}
\setcounter{equation}{0}

\subsection{Leptonic Asymmetries}\label{sec:las}

In leptogenesis, the CP violating effects that lead to the generation
of a net baryon asymmetry come from the difference between the decay
rate of heavy neutrinos into Higgs and leptons, and their
charge-conjugate processes. In RL models, the absorptive part of the
wavefunction contribution to the decay rate~\cite{PhysRevD.48.4609} is
central to capture the resonance effects that arise in models with
nearly degenerate singlet neutrino masses, and that result in the
enhancement of CP violation~\cite{Pilaftsis:1997dr}. To facilitate the
presentation of the analytic results for the CP asymmetry in heavy
neutrino decays within this framework, we introduce the
coefficients~\cite{Pilaftsis:2003gt,Pilaftsis:2005rv} 
\begin{align}
    A_{\alpha \beta} &= \sum_{l=1}^3 \frac{\h^\nu_{l \alpha}\h_{l
                       \beta}^{\nu*}}{16\pi} =
                       \frac{(\h^{\nu\dagger}\h^\nu)^*_{\alpha\beta}}{16\pi},\\ 
    V_{l \alpha} &= \sum_{k=1}^3 \sum_{\gamma \neq \alpha}
                   \frac{\h^{\nu*}_{k\alpha}\h^\nu_{k
                   \gamma}\h^\nu_{l\gamma}}{16\pi} f\left(
                   \frac{m^2_{N_\gamma}}{m^2_{N_\alpha}} \right), 
\end{align}
which pertain to the absorptive transition amplitudes for the
propagator and vertex, respectively. Here ${f(x) = \sqrt{x}\left[
    1-(1+x)\ln \left( \frac{1+x}{x} \right) \right]}$ is the
Fukugita-Yanagida one-loop function~\cite{FukYan:1986}. 

A full and consistent resummation of the CP-violating loop
corrections, including three Majorana neutrino mixing, generates the
following effective $L \tilde{\Phi} N$ Yukawa
couplings~\cite{Pilaftsis:2003gt,Pilaftsis:2005rv,Deppisch:2010fr}: 
\begin{align}
\label{eq:Eff_Yuk}
    (\bar{\mathbf{h}}^\nu_+)_{l\alpha} =&\; \h^\nu_{l\alpha} +
                                          iV_{l\alpha} - i
                                          \sum_{\beta,\gamma = 1}^3
                                          |\varepsilon_{\alpha\beta\gamma}|\,\h^\nu_{l\beta}\nonumber\\&\times
  \frac{m_{N_\alpha}\left(M_{\alpha\alpha\beta}+M_{\beta\beta\alpha}\right)-i
  R_{\alpha\gamma}
  \left[M_{\alpha\gamma\beta}\left(M_{\alpha\alpha\gamma}+M_{\gamma\gamma\alpha}\right)
  +
  M_{\beta\beta\gamma}\left(M_{\alpha\gamma\alpha}+M_{\gamma\alpha\gamma}\right)\right]}{m_{N_\alpha}^2-m_{N_\beta}^2
  + 2i m^2_{N_\alpha} A_{\beta\beta} + 2i\,\Im m
  R_{\alpha\gamma}\left(m_{N_\alpha}^2 |A_{\beta\gamma}|^2 +
  m_{N_\beta} m_{N_\gamma} \Re e A_{\beta\gamma}^2 \right)}\;, 
\end{align}
where $\epsilon_{\alpha\beta\gamma}$ is the anti-symmetric Levi-Civita
symbol, $M_{\alpha\beta\gamma}\equiv m_{N_\alpha}A_{\beta\gamma}$ and 
\begin{equation}
    R_{\alpha\beta} \equiv \frac{m_{N_\alpha}^2}{m_{N_\alpha}^2-m_{N_\beta}^2+2i m_{N_\alpha}^2 A_{\beta\beta}}\;.
\end{equation}
The corresponding CP-conjugate effective Yukawa coupling, which is
associated to the $L^C \tilde{\Phi}^* N$ interaction, is denoted by
$(\bar{\mathbf{h}}^\nu_-)_{l\alpha}$, and it can be found through the
replacement of $\h^\nu_{l\alpha}$ with $(\h^\nu_{l\alpha})^*$
in~\refs{eq:Eff_Yuk}. Notably, this resummed Yukawa coupling captures
all possible degrees of resonance between the contributions to the CP
asymmetry from the mixing between the singlet neutrinos, which
includes the \textit{bi-resonant} and \textit{tri-resonant}
cases. We should clarify here that the bi-resonant case implies
maximally enhanced CP asymmetries through the mixing of two singlet
neutrinos, and the tri-resonant implies maximally enhanced CP
asymmetries through the mixing of all three singlet
neutrinos. Moreover, in this formalism CP violation comes from the
difference between the resummed Yukawa couplings
$(\bar{\mathbf{h}}^\nu_-)_{l\alpha}$ and
$(\bar{\mathbf{h}}^\nu_+)_{l\alpha}$, as it can be seen by calculating
the heavy neutrino decay rates and scattering amplitudes with the help
of~\refs{eq:Eff_Yuk}. Note that for a model with two right-handed
neutrinos (or equivalently, for a model utilising the bi-resonant
approximation), the resummed Yukawa matrices are found by setting
$R_{\alpha\beta}$ to zero in~\refs{eq:Eff_Yuk}. 

Using these effective Yukawa couplings, the partial decay widths of the heavy neutrinos read
\begin{equation}
    \Gamma (N_\alpha \rightarrow L_l \Phi) =
    \frac{m_{N_\alpha}}{8\pi}\left| (\bar{\h}^\nu_+)_{l \alpha}
    \right|^2,\qquad 
    \Gamma (N_\alpha \rightarrow L^C_l \Phi^\dagger) =
    \frac{m_{N_\alpha}}{8\pi}\left| (\bar{\mathbf{h}}^\nu_-)_{l
        \alpha} \right|^2\;. 
    \label{eq:Gammas_def}
\end{equation}
In turn, these decay rates can be used to find the size of the CP
asymmetries for each lepton family, which for a given right-handed
neutrino $N_\alpha$ are defined as 
\begin{equation}
    \delta_{\alpha l} \equiv \frac{\Gamma (N_\alpha \rightarrow L_l
      \Phi) - \Gamma (N_\alpha \rightarrow L^C_l \Phi^\dagger) }{
      \sum_{k=e,\mu,\tau} \Gamma (N_\alpha \rightarrow L_k \Phi)
      +\Gamma (N_\alpha \rightarrow L^C_k \Phi^\dagger)} =
    \frac{\left| (\bar{\mathbf{h}}^\nu_+)_{l \alpha} \right|^2 -
      \left| (\bar{\mathbf{h}}^\nu_-)_{l \alpha}
      \right|^2}{(\bar{\mathbf{h}}^{\nu\dagger}_+\bar{\mathbf{h}}^\nu_+)_{\alpha\alpha}
      +
      (\bar{\mathbf{h}}^{\nu\dagger}_-\bar{\mathbf{h}}^\nu_-)_{\alpha\alpha}}. 
    \label{eq:CP-deltas_def}
\end{equation}
We also define the total CP asymmetry, $\delta_\alpha$, associated
with each heavy neutrino species: 
\begin{equation}
    \delta_\alpha \equiv \sum_{l=e,\mu,\tau} \delta_{\alpha l} \;.
    \label{eq:delta_N_def}
\end{equation}
In particular, a non-vanishing $\delta_\alpha$ may only be generated in
models, for which the flavour- and rephasing-invariant CP-odd quantity 
\begin{align}
    \Delta_{CP} &= \Im m \left\{ \textrm{Tr} \left[
                  (\mathbf{h}^\nu)^\dagger \mathbf{h}^\nu
                  \mathbf{m}_M^\dagger \mathbf{m}_M
                  \mathbf{m}_M^\dagger (\mathbf{h}^\nu)^{\sf T}
                  (\mathbf{h}^\nu)^* \mathbf{m}_M\right] \right\}\\ 
    &= \sum_{\alpha<\beta} m_{N_\alpha} m_{N_\beta}
      \left(m_{N_\alpha}^2 - m_{N_\beta}^2 \right)\, \Im m \Big[
      \left(\mathbf{h}^{\nu\dagger}\mathbf{h}^\nu
      \right)_{\beta\alpha}^2 \Big]
\end{align}
is non-zero~\cite{Pilaftsis:1997jf,Pilaftsis:2003gt,Branco:1986gr,Yu:2020gre}. For the model
presented in Section~\ref{sec:Z_6_seesaw}, this CP-odd quantity may be
expressed as 
\begin{align}
    \Delta_{CP} &\approx \left(a^2 + b^2 + c^2 \right)^2
                  \sum_{\alpha<\beta} m_{N_\alpha} m_{N_\beta}
                  \left(m_{N_\alpha}^2 - m_{N_\beta}^2 \right)\, \Im m
                  \Big( \omega^{2(\alpha-\beta)} \Big)\; . 
\end{align}
When all heavy neutrino masses are exactly degenerate, the CP-odd invariant
$\Delta_{CP}$ vanishes. However, with the inclusion of mass differences,
$\Delta_{CP}$ is proportional to the imaginary part of the
$\mathbb{Z}_6$ element $\omega^2$ only.
\begin{figure}[t!]
\hspace*{-0.6cm}
    \includegraphics[width=17.5cm]{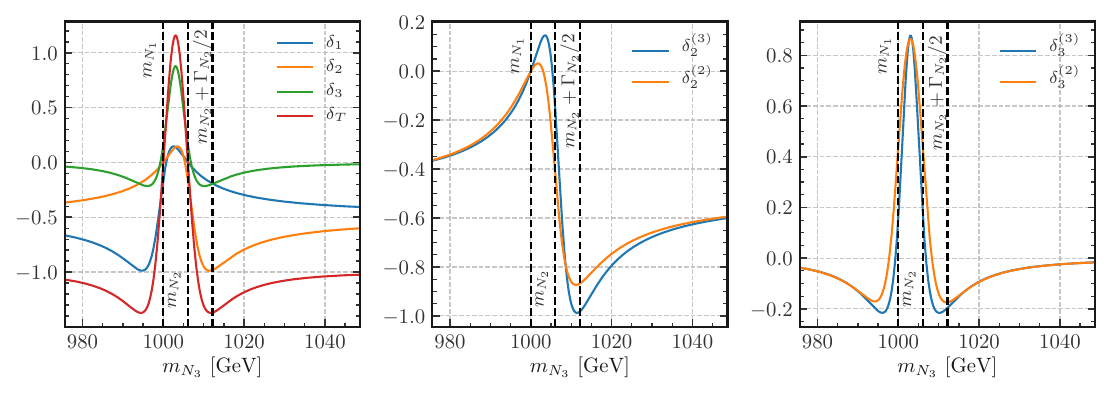}
    \caption{\textit{Left panel:} CP asymmetries in the decays of
      $N_1$, $N_2$ and $N_3$, together with the total CP asymmetry
      $\delta_T = \sum_\alpha \delta_\alpha$, as a function of the
      mass of $N_3$. \textit{Centre panel:} CP asymmetry in the decay
      of $N_2$ vs. $m_{N_3}$ as calculated in a model that considers
      two-neutrino mixing ($\delta^{(2)}_2$) and three-neutrino mixing
      ($\delta^{(3)}_2$). \textit{Right panel:} CP asymmetry in the
      decay of $N_3$ vs. $m_{N_3}$ calculated in a model that
      considers two-neutrino mixing ($\delta^{(2)}_3$) and
      three-neutrino mixing ($\delta^{(3)}_3$). In all three panels,
      the vertical dashed lines indicate, from left to right, the
      values of $m_{N_1}$, $m_{N_2}$ and the tri-resonant value of
      $m_{N_3}$ (for details, see text).} 
    \label{fig:cp_asymmetry}
\end{figure}

Several applications of the RL formalism
(e.g.~\cite{Gu:2010xc,Garny:2011hg,Aoki:2015owa,Asaka:2018hyk,GRANELLI2021115597,Chauhan:2021xus,CHAKRABORTY2022115780})
exploit the bi-resonant enhancement of CP violating effects due to the
mixing of two Majorana neutrinos, while the contribution to the CP
asymmetry due a third singlet neutrino is either absent due to the
neutrino mass model choice, or negligible when compared to the one
generated in the decays of the resonating pair. However, in a model
with three right-handed neutrinos, in the region where the masses of
the heavy neutrinos satisfy the resonance condition 
\begin{equation}
|m_{N_\alpha}-m_{N_\beta}|\simeq\frac{\Gamma_{N_{\alpha,\beta}}}{2}\qquad(\alpha\neq\beta),\label{eq:resonant_condition}
\end{equation}
effects of constructive interference generated by a third resonating
neutrino can further enhance CP violation as compared to the case when
only two neutrinos are in resonance. Figure~\ref{fig:cp_asymmetry}
shows the behaviour of the CP asymmetries in the decays of $N_1$,
$N_2$, and $N_3$, as well as the total CP asymmetry
$\delta_T=\sum_\alpha\delta_\alpha$, plotted against $m_{N_3}$. In
this figure, the mass of $N_2$ is fixed at the value
$m_{N_2}=m_{N_1} + \Gamma_{N_1}/2$, therefore it fulfills the
bi-resonant condition. On the left panel, it can be seen that when
$m_{N_3}=m_{N_2}$, the total CP asymmetry (solid red line) vanishes
due to the destructive interference effect of $N_3$, while at
$m_{N_3}=m_{N_2}+\Gamma_{N_2}/2$, $|\delta_T|$ reaches a maximum that
is more than 35\% higher than in a model where the mass of the third
singlet neutrino lies outside the resonance region (i.e., high
$m_{N_3}$). In this tri-resonant point, one has
$\delta_1\approx\delta_3$, while $\delta_2$ is the dominant
contribution to $\delta_T$. Furthermore, we find that the values of
$\delta_{1,2,3}$ are independent of the mass scale, $m_{N_1}$,
provided that the tri-resonant condition is satisfied. Thus, the
enhancement of $\delta_2$ is pervasive throughout the tri-resonant
parameter space. The middle panel of Figure~\ref{fig:cp_asymmetry}
shows the impact of the proper three-neutrino mixing resummation on
the asymmetry $\delta_2$ by comparing the asymmetry calculated by
considering three Majorana neutrino mixing ($\delta^{(3)}_2$) with the
two-neutrino mixing case ($\delta^{(2)}_2$). When $m_{N_3}$ lies in
the resonance region, it can be seen that the mixing with $N_3$
becomes important, and there is a sizeable difference between the two-
and three-neutrino mixing scenarios, where the latter has a sizeable
enhancement effect on $\delta_2$. The right panel of
Figure~\ref{fig:cp_asymmetry} shows that the inclusion of
three-neutrino mixing also affects the size of the maximum magnitude
of the asymmetry in the decays of $N_3$, although to a lesser extent.

Overall, Figure~\ref{fig:cp_asymmetry} showcases a resonant
enhancement of the total CP asymmetry of the model when the three
heavy neutrinos are in successive resonance, a scenario that we have
described as tri-resonant, in contrast to the bi-resonant
approximation commonly studied in the literature.  We identify a
particular tri-resonant structure which generates appreciable BAU and
maximises the scale of CP asymmetry within a model with three singlet
neutrino mixing. In the literature, there also exist studies which
consider the mixing effects of three singlet
neutrinos~\cite{Deppisch:2010fr,Pilaftsis:2015bja,Abada:2018oly,Drewes:2021nqr,Drewes:2022kap,Granelli:2022eru}.
  These studies utilise a flavour structure different to the
  $\mathbb{Z}_6$ structure we have adopted, and in the case of
  \cite{Drewes:2021nqr}, it is more similar to that proposed in
  \cite{Pilaftsis:2004xx}. Hence, the flavour structure presented in these
  studies cannot be mapped onto the discrete flavour symmetries we
  have used here, so as to enable some meaningful comparison.
Finally, we must point out  that our approximate $\mathbb{Z}_6$-symmetric
flavour structure provides both light neutrino masses, and the origin for CP violation.

\subsection{Boltzmann Equations}\label{sec:bes}
The conditions for generating a BAU, dictated
by~\cite{Sakharov:1967dj}, require not only a violation of the CP
symmetry, but also a departure from thermal equilibrium and baryon
number violation. Here, we introduce the set of Boltzmann equations
that describe the out-of-equilibrium dynamical generation of a lepton
asymmetry in the early Universe, and assume that it is reprocessed
into a net baryon number through equilibrium $(B +L)$-violating
sphaleron transitions~\cite{KUZMIN198536}. 

At temperatures, $T$, pertinent to leptogenesis, the Universe is assumed to be radiation dominated, with an energy and entropy density given by
\begin{align}
    \rho(T) &=\frac{\pi^2}{30} \geff(T) \, T^4 \;,
    \label{eq:rho_def} \\
    s(T) &= \frac{2\pi^2}{45} \heff(T) \, T^3 \;,
    \label{eq:s_def}
\end{align}
respectively. Here $\geff$ and $\heff$ are the relativistic dofs of
the SM plasma that correspond to $\rho$ and $s$, respectively. For our
numerical results, we use the tabulated data\footnote{We have
  extracted the corresponding data file from the source code of {\tt
    MicrOMEGAs}~\cite{Belanger:2013oya}.} for the relativistic dofs as
calculated in~\cite{Hindmarsh:2005ix}.\footnote{From
  \cite{Hindmarsh:2005ix} we choose the equation of state model
  labeled as C.} 

The evolution of the heavy neutrino and lepton asymmetry number
densities are described by their respective BEs in terms of the
dimensionless parameter $\za = m_{N_{\alpha}}/T$, for
$\alpha=1,2,3$. In line with previous conventions, we use
$z=z_1$. These BEs are presented in~\cite{Pilaftsis:2005rv}, and due
to the approximate democratic structure of the neutrino Yukawa matrix in our TRL
models, we sum over lepton flavours, which leaves us with four coupled
evolution equations.  

Following the conventions of~\cite{Pilaftsis:2005rv}, we normalise all
number densities with the photon number density 
\begin{equation}
    n_{\gamma}(\za) = \frac{2\zeta(3)T^3}{\pi^2} = \frac{2\zeta(3)}{\pi^2} \lrb{\dfrac{\mNa}{\za}}^3,
\end{equation}
which for a given particle species $i$, gives us the ratio
\begin{align}
    \eta_{i}(\za) &= \frac{n_i(\za)}{n_{\gamma}(\za)}\;.
    \label{eq:eta_def}
\end{align}
In addition, we define the departure from equilibrium for the heavy-neutrino density as
\begin{align}
    \dNa(\za) &= \frac{\etaNa(\za)}{\etaNeq(\za)} -1 \;,
    \label{eq:dNa_def} 
\end{align}
where $\etaNeq$ denotes $\etaNa$ in thermal equilibrium, for which we
use the approximate expression 
\begin{equation}
    \etaNeq(\za) \approx \frac{\za^2}{2\zeta(3)}K_2(\za).
\end{equation}
Here, $\zeta(3)\approx 1.202$ is Ap\'ery's constant, and $K_n(z)$ is a
modified Bessel function of the second kind. In the BEs, we have also
included terms which depend on the parameter 
\begin{equation}
    \delta_h (\za) = 1 - \dfrac{1}{3} \dfrac{d \ln \heff }{d \ln \za}\;,
    \label{eq:deltah_def}
\end{equation}
since we allow $h_\textrm{eff}$ to vary with $T$.\footnote{In fact, in
  the data file we have extracted from {\tt MicrOMEGAs}, the
  relativistic dofs are not constant even at  temperatures well above
  $100~\GeV$.  This unexpected behaviour arises from combined lattice \cite{Karsch:2000ps} and perturbative QCD \cite{Kajantie:2002wa}
  considerations to the equation of
  state of the plasma, leading to 
deviations from the ideal gas assumption at high temperatures \cite{Hindmarsh:2005ix}.} 

Considering decay terms, $\Delta L=1$ and $\Delta L=2$ scattering
processes, and the running of the dof parameters, the BEs can be
written as\footnote{In order to solve the system of
  equations~\refs{eq:BEN,eq:BEL}, we employ the implementation of {\tt
    RODASPR2}~\cite{RANG2015128} provided in {\tt
    NaBBODES}~\cite{NaBBODES}. We have checked that other
  methods~\cite{Rentrop1979,RangAngermann2005} as well as the ones
  provided by {\tt scipy}~\cite{2020SciPy-NMeth} produce the same
  results. The relativistic dofs of the plasma are interpolated using
  {\tt SimpleSplines}~\cite{SimpleSplines} and the various integrals
  needed for the collision terms are evaluated using {\tt
    LAInt}~\cite{LAInt}. Finally, all figures are made using the
  versatile visualization library {\tt
    matplotlib}~\cite{Hunter:2007}. } 
\allowdisplaybreaks
\begin{align}
    \frac{d\dNa}{d \ln \za} =&-\dfrac{ \delta_h(\za)}{H(\za) \ \eta_{N_\alpha}^{\rm eq}(\za) }\lrsb{\dNa  \lrb{\Gamma^{D(\alpha)} + \Gamma^{S(\alpha)}_Y + \Gamma^{S(\alpha)}_G} +\frac{2}{9}\, \etaL\, \delta_\alpha \lrb{\tilde\Gamma^{D(\alpha)} + \hat{\Gamma}^{S(\alpha)}_Y + \hat{\Gamma}^{S(\alpha)}_G} } \nonumber\\
    & +\lrb{\dNa+1} \, \left[\za \frac{K_1(\za)}{K_2(\za)} - 3(\delta_h(\za) -1) \right]\;,
    \label{eq:BEN}\\
    \frac{d\eta_L}{d \ln z} =& -  \frac{\delta_h(z)}{H(z)} \lrBiggcb {\sum_{\alpha=1}^3 \dNa \delta_\alpha \lrb{ \Gamma^{D(\alpha)} + \Gamma^{S(\alpha)}_Y + \Gamma^{S(\alpha)}_G} \nonumber \\
    & +\frac{2}{9}\eta_L \left[\sum_{\alpha=1}^3\left(\tilde\Gamma^{D(\alpha)} + \tilde{\Gamma}^{S(\alpha)}_Y + \tilde\Gamma^{S(\alpha)}_G + 
    \Gamma^{W(\alpha)}_Y + \Gamma^{W(\alpha)}_G\right) + \Gamma^{\Delta L = 2} \right] \nn\\ 
    & + \dfrac{2}{27} \etaL \, \sum_{\alpha=1}^3 \delta_\alpha^2 \, \lrb{ \Gamma^{W(\alpha)}_Y + \Gamma^{W(\alpha)}_G } }
    - 3\etaL(\delta_h(z) -1) \label{eq:BEL}\;,
\end{align}
where
\begin{equation}
    H(\za) = \sqrt{\frac{4\pi^3 g_{\textrm{eff}}(\za)}{45}}\frac{m_{N_\alpha}^2}{M_{\textrm{Pl}}} \frac{1}{\za^2}
\end{equation}
is the Hubble parameter, and
$M_{\textrm{Pl}} \approx 1.221 \times 10^{19} ~\GeV$ is the Planck
mass. Since the BEs are not identical to the ones utilised
in the literature due to the non-trivial $T$-dependence of
$h_{\rm eff}$, we show how they are obtained in
Appendix~\ref{app:BE_dh}. The various collision terms are defined in
the literature~\cite{Pilaftsis:2005rv} as \allowdisplaybreaks
\begin{align}
    \Gamma^{D(\alpha)} &= \frac{1}{n_\gamma}\gamma^{N_\alpha}_{L\Phi},\\
    \tilde\Gamma^{D(\alpha)} &= \lrb{1+\dfrac{12}{21}} \Gamma^{D(\alpha)},\\
    \Gamma^{S(\alpha)}_Y &= \frac{1}{n_\gamma}\left[\gamma^{N_\alpha L}_{Qu^C} + 2\gamma^{N_\alpha u^C}_{LQ^C} \right],\\
    \tilde{\Gamma}^{S(\alpha)}_Y &= \frac{1}{n_\gamma}   
    \lrsb{ \lrb{\dNa +1 + \dfrac{12}{21}} \gamma^{N_\alpha L}_{Qu^C} + \lrb{2 + \dfrac{98}{159}(\dNa+2) } \gamma^{N_\alpha u^C}_{LQ^C} },\\
    \hat{\Gamma}^{S(\alpha)}_Y &= \frac{1}{n_\gamma}   
    \lrsb{ \lrb{-(\dNa +1) + \dfrac{12}{21}} \gamma^{N_\alpha L}_{Qu^C} + \lrb{2 - \dfrac{98}{159}\dNa } \gamma^{N_\alpha u^C}_{LQ^C} },\\
    \Gamma^{S(\alpha)}_G &= \frac{1}{n_\gamma}\left[\gamma^{N_\alpha V_\mu}_{L\Phi} +\gamma^{N_\alpha L}_{V_\mu \Phi^\dagger} +\gamma^{N_\alpha \Phi^\dagger}_{L V_\mu} \right],\\
    \tilde{\Gamma}^{S(\alpha)}_G &= \frac{1}{n_\gamma}\lrsb{
    \lrb{ 1+\dfrac{12}{21} }\gamma^{N_\alpha V_\mu}_{L\Phi}
    +\lrb{ \dNa + 1+\dfrac{12}{21} } \gamma^{N_\alpha L}_{V_\mu \Phi^\dagger}
    +\lrb{1+(\dNa+1)\dfrac{12}{21}}\gamma^{N_\alpha \Phi^\dagger}_{L V_\mu} },\\
    \hat{\Gamma}^{S(\alpha)}_G &= \frac{1}{n_\gamma}\lrsb{
    \lrb{ 1+\dfrac{12}{21} }\gamma^{N_\alpha V_\mu}_{L\Phi}
    +\lrb{ -(\dNa + 1)+\dfrac{12}{21} } \gamma^{N_\alpha L}_{V_\mu \Phi^\dagger}
    +\lrb{1-(\dNa+1)\dfrac{12}{21}}\gamma^{N_\alpha \Phi^\dagger}_{L V_\mu} },\\
    \Gamma^{W(\alpha)}_Y &= \frac{1}{n_\gamma}\lrsb{ 
    \lrb{2+\dfrac{12}{21}} \gamma^{N_\alpha L}_{Qu^C} + \lrb{2+\dfrac{12}{7}} \gamma^{N_\alpha u^C}_{LQ^C}},\\
    \Gamma^{W(\alpha)}_G &= \frac{1}{n_\gamma} \lrsb{
    \lrb{1+\dfrac{12}{21}}\gamma^{N_\alpha V_\mu}_{L\Phi} + 
    \lrb{2+\dfrac{12}{21}}\gamma^{N_\alpha L}_{V_\mu \Phi^\dagger} +
    \lrb{1+\dfrac{24}{21}}\gamma^{N_\alpha \Phi^\dagger}_{L V_\mu} },\\
    \Gamma^{\Delta L=2} &= \frac{2}{n_\gamma}\lrb{1+\dfrac{12}{21}}\left[ {\gamma'}^{L \Phi}_{L^C \Phi^\dagger} +\gamma^{LL}_{\Phi^\dagger\Phi^\dagger} \right],
\label{eq:rates}
\end{align}
where $\gamma^X_Y$ are CP-conserving collision terms for the process ${X\rightarrow Y}$. The latter is defined as
\begin{align}
\gamma^X_Y\equiv\gamma(X\rightarrow Y) + \gamma(\overline{X}\rightarrow \overline{Y}),
\end{align}
where the bar denotes CP conjugation. The pertinent analytical expressions of the collision terms and scattering cross sections can all be found in~\cite{Pilaftsis:2005rv}.\footnote{For the gauge and Yukawa mediated cross section, we use the lepton thermal mass, as infra-red regulator~\cite{Weldon:1982bn}.} Note that the primed terms correspond to collision terms with subtracted real intermediate states (RIS), which can take negative values due to the lack of an on-shell contribution to the squared amplitude. 
\begin{figure}[t!]
    \centering
    \includegraphics[width=0.6\textwidth]{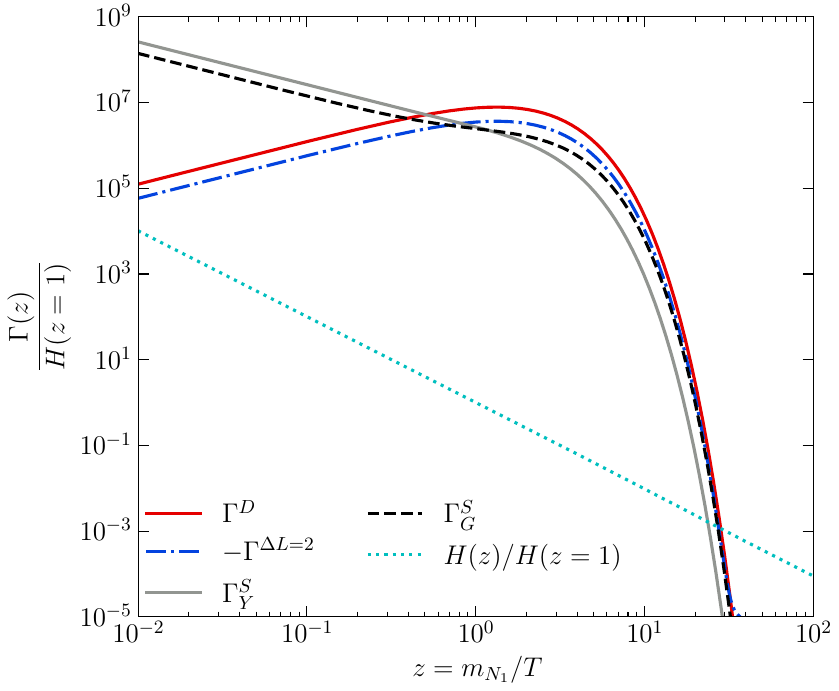}
    \caption{The $\dNa$ independent collision terms are defined in (\ref{eq:rates}) for $|\h^{\nu}_{ij}| \approx 3 \times 10^{-3} $ and $m_{N_{1}} = 500~\GeV$. The wash-out terms, $\Gamma^W_{Y,G}$, are uniformly different by a factor of two compared to their source ($\Gamma^S_{Y,G}$) counterparts.}
    \label{fig:rates}
\end{figure}

The typical dependence of the various collision terms on $z = m_{N_1}/T$ is
shown in Figure~\ref{fig:rates}, for
$|\h^{\nu}_{ij}|\approx3\times10^{-3}$ and $m_{N_1} = 500~\GeV$.  The
other two masses obey the tri-resonant condition, which results in a
sizeable $\Delta L = 2$ rate. It is noteworthy that the collision term
that describes the decays and the RIS parts is larger than
$\Gamma^{\Delta L = 2}$, as also observed
in~\cite{Pilaftsis:2003gt}. For this figure, the relevant perturbation
matrix, $\delta \h^\nu$, needed to match the neutrino data may be
found in Appendix~\ref{app:deltah_values} under Benchmark \textbf{A}. 

During leptogenesis, part of the lepton asymmetry that is generated in
the processes described above is partially converted into a baryon
asymmetry by $(B+L)$-violating sphaleron transitions which become
exponentially suppressed below the temperature  $\Tsph\simeq 132$
GeV~\cite{DOnofrio:2014rug}. In order to compare the generated BAU at
$T=\Tsph$ to its value at the recombination epoch, we assume that
there are no considerable entropy releasing processes, and hence the
entropy density remains approximately constant as the Universe
cools. Using entropy conservation and the relation ${s(T) \sim
  h_{\textrm{eff}}(T)\,  T^3}$, it can be shown that the BAU at
$T_\textrm{sph}$ is related to the BAU at $T_\textrm{rec}$ by 
\begin{align}
    \eta^{\rm rec}_B=\frac{\heff(T_{\rm rec})}{\heff(\Tsph)}\frac{n_B(\Tsph)}{n_\gamma(\Tsph)}=f\frac{n_B(\Tsph)}{n_\gamma(\Tsph)} \;.
    \label{eq:etaB_obs_def}
\end{align}
For the dilution factor, $f$, we use the approximate value
$1/27$~\cite{Buchmuller:2004nz,Pilaftsis:2003gt}, while for the
conversion factor between lepton and baryon number above the sphaleron
temperature, we use the equilibrium relation given
by~\cite{PhysRevD.42.3344} 
\begin{align}
\eta_B = -\frac{28}{51}\eta_L \;.
    \label{eq:etaL_to_etaB}
\end{align}

\section{Approximate Solutions to Boltzmann Equations}\label{sec:approx}
\setcounter{equation}{0}
In this section we discuss the solution of the BEs~\eqs{eq:BEN,eq:BEL}
in order to understand the production of a lepton asymmetry in the
early Universe. As a first approach, we consider a simplified version
of these equations, where we ignore the ``back-reaction" (\ie~the
second term of~\refs{eq:BEN}), the variation of the relativistic dofs,
and only take into account the decay and RIS terms. Moreover, we
assume that $m_{N_1} \approx m_{N_2} \approx m_{N_3}$. 

\subsection{Approximation for \texorpdfstring{$\dNa$}{dNa}}  
We begin by solving the equation for $\dNa$, which takes the form
\begin{align}
    &\dfrac{d \dNa}{d z } = \dfrac{K_1(z)}{K_2(z)} \lrsb{1 + 
    \lrb{1- z\dfrac{\GNa}{ \HN } }  \  \dNa}  \;.
    \label{eq:BEN_approx}
\end{align}
Initially (at $z \ll 1$), right-handed neutrinos are taken to be in thermal
equilibrium, so $\dNa = 0$. Therefore, at early times, we expect the
second term of~\refs{eq:BEN_approx} to vanish. Moreover, at such high
temperatures, we may approximate ${K_1(z) / K_2 (z) \approx z/2}$, so 
\begin{equation}
    \dNa \approx \dfrac{z^2}{4} \,, \qquad \text{for } z \ll 1 \;.
    \label{eq:dNa_approx_init}
\end{equation}
As the temperature drops, $\dNa$ increases, and at some point the
second term starts to become comparable to the first. So, $\dNa$
continues to increase until  
both terms become equal. We denote this point as $z=\hat z$, and
assuming $\hat z \gg \HN / \GNa$, it is estimated as 
\begin{equation}
    \hat z \approx  
    \lrb{\dfrac{4 \, \HN }{\GNa }}^{1/3}\;.
    \label{eq:zNmax}
\end{equation}
For $z \approx \hat z$, we observe that the \rhs
of~\refs{eq:BEN_approx} stays close to zero. That is, $\dNa \approx
\HN / \GNa z^{-1}$, since any increase (decrease) with respect to this
behaviour pushes $\dNa$ to negative (positive) values. Consequently, we
find that for $z \gg \hat z$,  
\begin{equation}
    \dNa \approx \dfrac{\HN}{ \GNa  z}\;.
    \label{eq:dNa_approx_late}
\end{equation}
Notice that this result does not depend on the initial
condition. Also, we should point out that at late times, namely $z \gg
1$,~\refs{eq:dNa_approx_late} solves~\refs{eq:BEN_approx} up to terms
$\mathcal{O}(1/z^2)$. 

\subsubsection{The Neutrino Boltzmann Equation as an Autonomous System}
\begin{figure}[t!]
    \centering
    \includegraphics[width=1.\textwidth]{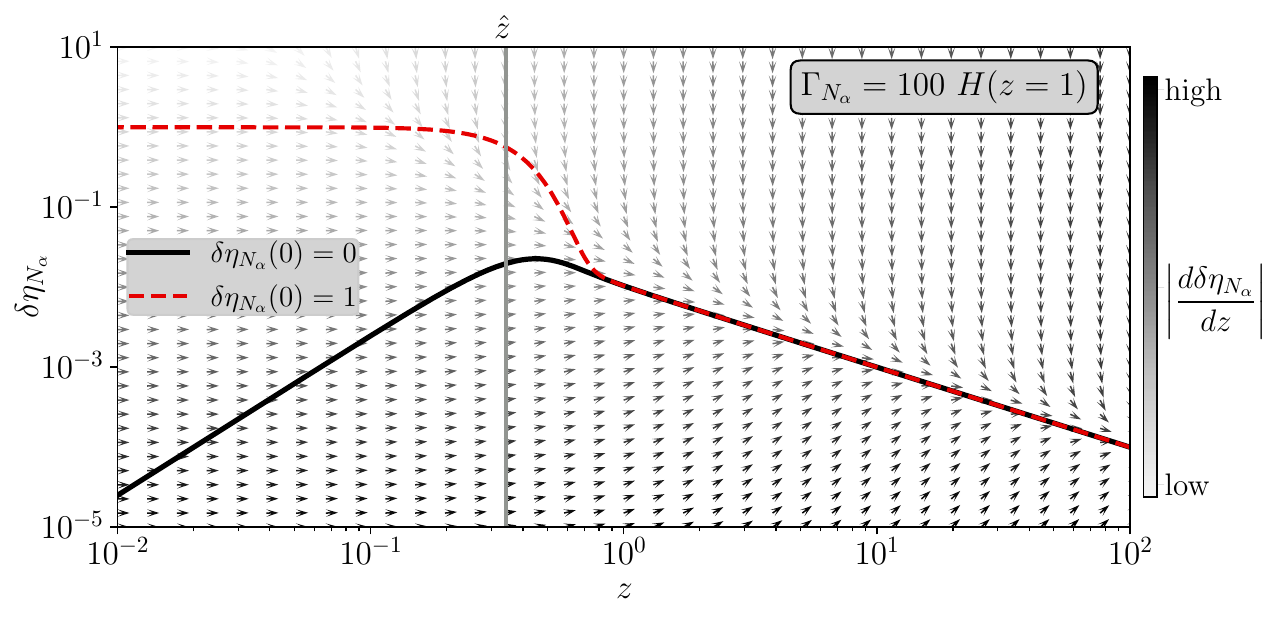}
    \caption{The solution of (\ref{eq:BEN_approx}) for $\GNa=100 \
      \HN$ with initial conditions $\dNa(z \to 0)=0$ (black) and
      $\dNa(z\to 0)=1$ (dashed red). The arrows show the direction of
      $\mathbf{V}$, while the colour gradient encodes the size of
      $\dNa$, with light grey (black) for low (high) values of
      $\left|d\dNa/dz\right|$. The vertical grey line shows the value
      of $\hat z$ as estimated by (\ref{eq:zNmax}).} 
    \label{fig:BEN_flow}
\end{figure}

The independence from the initial conditions has been previously
highlighted in the literature (\eg
~\cite{Pilaftsis:2005rv,Deppisch:2010fr}). However, it would be
helpful to analyse its attractor properties. We begin by noting
that~\refs{eq:BEN_approx} can be written in the form of an autonomous
system  
\begin{equation}
    \dfrac{d \mathbf{r}}{d t} = \mathbf{V}\lrb{z(t),\dNa(t)} \;,
    \label{eq:autonomous_BEN}
\end{equation}
with $\mathbf{r} = (z,\dNa)^{\sf T}$ and
\begin{equation}
    \mathbf{V}\lrb{z(t),\dNa(t)} = \lrb{
    \begin{matrix}
    1 \\[0.3cm]
    \dfrac{K_1(z(t))}{K_2(z(t))} \lrsb{1+\lrb{1- z(t) \dfrac{\GNa}{ \HN } }  \  \dNa(t) }   
    \end{matrix}
    }.
    \label{eq:velocity_BEN}
\end{equation}
Here, the vector field $\mathbf{V}$ represents the flow of~\refs{eq:BEN_approx}, which helps to demonstrate how $\mathbf{r}$ reaches the stable solution, independently of the initial conditions. 
In Figure~\ref{fig:BEN_flow}, we show the evolution of $\dNa$ for
$\GNa=100\, \HN$ and for two different initial conditions. Along with
the two curves, we show the direction of $\mathbf{V}$, which indicates
at each point the tendency of $\mathbf{r}$. Moreover, darker arrows
imply higher values of $\left| d\dNa/dz \right|$.   
As both curves merge at $z \gtrsim \hat z$,  $\dNa$ ends up becoming
ignorant of the initial condition.  
This feature is also imprinted in the direction of~$\mathbf{V}$. The
normalised vector, $\mathbf{V}$, is parallel to the $z$-axis  for $z
\lesssim \hat z$, while it points towards the solution for $z \gtrsim
\hat z$. 

\subsection{Approximation for \texorpdfstring{$\etaL$}{etaL}}
The corresponding equation for the lepton asymmetry, assuming that $\dNa \sim 1/z$, can be written as
\begin{align}
    &\dfrac{d \etaL}{d z} = \dfrac{\delta_T}{2\zeta(3)} K_1(z) \, z^2 \ \lrb{1-z \, \dfrac{2 k_L}{3 \delta_T} \etaL} \;, 
    \label{eq:BEL_approx}
\end{align}
where $\delta_T = \sum_\alpha \delta_\alpha$  and $k_L = \displaystyle\sum_a \dfrac{\GNa}{ \HN }$. 

Initially, at $z \ll 1$, the lepton asymmetry is assumed to vanish. So, at high temperatures, only the first term of~\refs{eq:BEL_approx} contributes. Therefore, since $K_1(z)z^2 \approx z$, $\etaL \approx  \delta_T/\lrb{4\zeta(3)} \, z^2$. As $\etaL$ increases, both terms become comparable at 
\begin{equation}
    z=\tilde z \approx 2 k_L^{-1/3}\;.
    \label{eq:ztilde_def}
\end{equation}
After this point, the \rhs of~\refs{eq:BEL_approx} remains close to zero, as in the previous case. Hence,
\begin{align}
    &\etaL \approx \dfrac{3\delta_T}{2 k_L\, z } \;. 
    \label{eq:etaL_brfore_fo}
\end{align}
However, at very low temperatures, $z \gg 1$, the \rhs of~\refs{eq:BEL_approx} becomes exponentially suppressed, due to the asymptotic behaviour $K_1(z \ll 1) \sim e^{-z}$. Then, the lepton asymmetry becomes a constant, \ie $\etaL$ freezes out at some $z=z_{\rm fo}$. This point can be estimated by demanding that the rate at which $\etaL$ changes is comparable to its magnitude, which implies that
\begin{equation}
    \dfrac{ k_{L} }{3\zeta(3)} \sqrt{\dfrac{\pi}{2}} e^{-z_{\rm fo}} \ z_{\rm fo}^{5/2} \approx 1 \;.
    \label{eq:zfo_def}
\end{equation}
This equation can be solved using fixed point iteration. Keeping the first two iterations, we estimate\footnote{This form agrees with our numerical solution of~\refs{eq:zfo_def} within $10 \%$.}
\begin{equation}
     z_{\rm fo} \approx  \ln(3k_{L}) + 5/2 \ln(\ln (3k_{L}))+ \mathcal{O}\lrb{\ln\ln\ln k_{L}} \;.
    \label{eq:z3}
\end{equation}
At lower temperatures, with $z>z_{\rm fo}$, the lepton asymmetry becomes 
\begin{align}
    &\etaL \approx \dfrac{3\delta_T}{2 k_L\, z_{\rm fo}} \;. 
    \label{eq:etaL_fo}
\end{align}
As before, we note that this solution is independent of any initial asymmetry that might have existed before the one generated by the heavy Majorana neutrino decays. 

\subsubsection{The Baryon Asymmetry}
Using~\refs{eq:etaB_obs_def,eq:etaL_to_etaB} and assuming that the freeze-out happens before $z=\zsph$, \ie $z_{\rm fo}>\mNa/\Tsph $, the predicted baryon asymmetry at the time of recombination reads
\begin{equation}
    \eta_B \sim - 3 \times 10^{-2} \ 
    \displaystyle  \dfrac{ \delta_T}{ k_{L} \left[\ln(3 k_{L}) + 5/2  \ln(\ln(3 k_{L}))  \right]} \;.    
    \label{eq:eta_B}
\end{equation}
However, if $z_{\rm fo}<\mNa/\Tsph$, the baryon asymmetry becomes
\begin{equation}
    \eta_B \sim - 3 \times 10^{-2} \ 
   \displaystyle \dfrac{ \delta_T}{ k_{L} \zsph  } \;.  
   \label{eq:eta_B_zc}
\end{equation}
It should be noted that the resulting value of $\etaL$ is proportional to $\delta_T / [k_{L} \ln(3 k_{L})]$, with the logarithmic dependence coming from the determination of the freeze-out temperature. This means that, generally, for a given value of $k_{L}$, only $\delta_T$ determines the baryon asymmetry. In particular, for  $k_{L}>10$, the observed baryon asymmetry can be obtained for $\delta_T \gtrsim 10^{-7}$. 

\subsubsection{The Lepton Asymmetry Boltzmann Equation as an Autonomous System} 
\begin{figure}[t!]
    \centering
    \includegraphics[width=1.\textwidth]{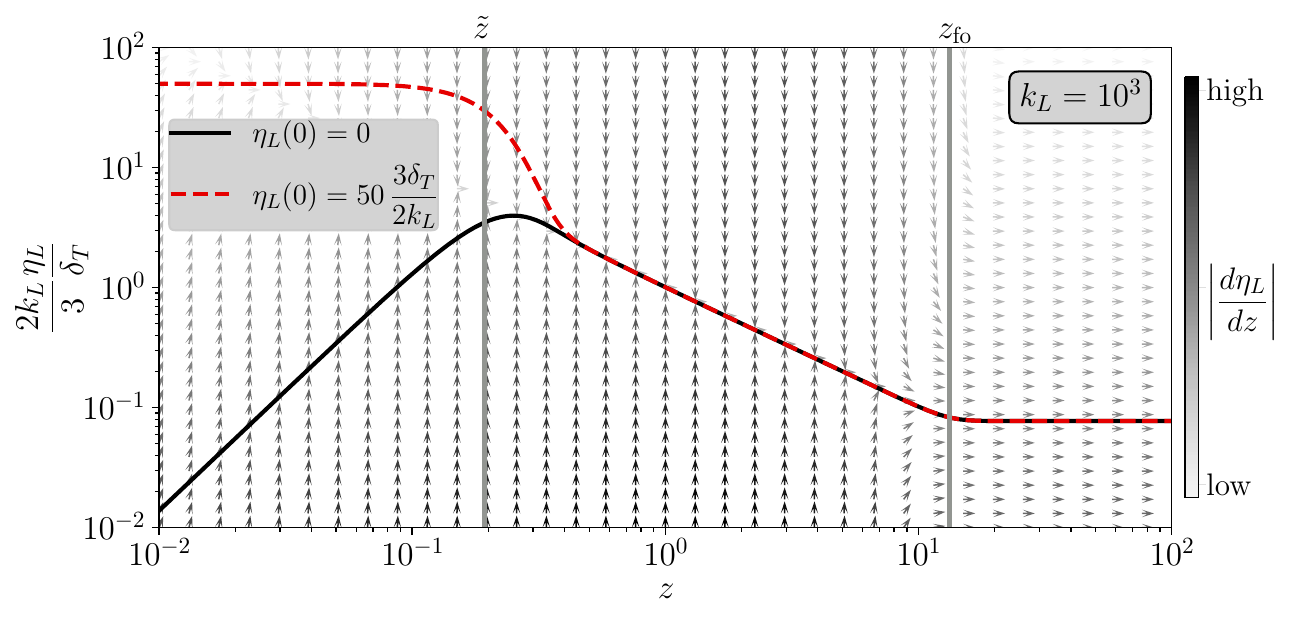}
    \caption{The solution of~(\ref{eq:BEL_approx}) for $k_{L}=100$ with initial conditions $\etaL(z \to 0)=0$ (black) and $\etaL(z \to 0)=50 \, (3\delta_T/2k_L)$ (dashed red). The arrows show the direction of $\mathbf{V}$, while the colour gradient encodes the size of the \lhs relative to $\etaL$; from light grey for low values to darker grey for higher values. The two vertical (grey) lines show the values of $z = \tilde z$ and $z = z_{\rm fo}$ as approximated in~(\ref{eq:ztilde_def}) and~(\ref{eq:z3}), respectively.}
    \label{fig:BEL_flow}
\end{figure}

The lepton asymmetry BE~(\ref{eq:BEL_approx}) can also be written as an autonomous system
\begin{equation}
     \dfrac{d \mathbf{r}}{d t} = \mathbf{V}\lrBigb{z(t),\etaL(t)} \;,
    \label{eq:auto_BEL}
\end{equation}
with $\mathbf{r} = (z,\etaL)^\textsf{T}$ and
\begin{equation}
    \mathbf{V}\lrBigb{z(t),\eta_{L}(t)} = \lrb{
    \begin{matrix}
    1 \\[0.3cm]
    \dfrac{\delta_T}{2\zeta(3)} K_1(z(t)) \, z^2(t) \ \lrb{1-z(t) \, \dfrac{2 k_L}{3 \delta_T} \etaL(t)}
    \end{matrix}
    } \;.
    \label{eq:velocity_BEL}
\end{equation}
This system is similar to~\refs{eq:velocity_BEN}, but it also exhibits a freeze-out. We can observe the evolution of $\etaL$ in Figure~\ref{fig:BEL_flow}, where we show the flow of the autonomous system of~\refs{eq:auto_BEL}, along with its solution for two initial conditions.
We note that the derivative of $\etaL$ can only deviate significantly from zero in the range $\tilde z \lesssim z \lesssim z_{\rm fo}$. This is the period in which $d\etaL/dz$ pushes the solution towards $\etaL \sim 1/z$, as below (above) this curve $\mathbf{V}$ points upwards (downwards), with significant magnitude. Notice that in the region $z\gtrsim z_{\rm fo}$, $\mathbf{V}$ points towards the right. This means that the component that dominates $\mathbf{V}$ is $d z / d t = 1$. Thus, the flow of~\refs{eq:auto_BEL} will only follow $z$, as $d\etaL/dz$ gets exponentially suppressed.

\subsection{Numerical Approximation of the Complete Boltzmann\\ Equations}

Although~\refs{eq:BEN_approx,eq:BEL_approx} are very different than
their more accurate counterparts, i.e.~the BEs~\refs{eq:BEN,eq:BEL},
they show that their solutions mostly follow the lines that cause the
\rhs to approximately vanish. In other words, they follow ``attractor"
solutions. The same argument used to show this
for~\refs{eq:BEN_approx,eq:BEL_approx} can be applied
to~\refs{eq:BEN,eq:BEL}. Assuming that the $\etaL$-dependent term
of~\refs{eq:BEN} is suppressed (\ie $\delta_\alpha \ll 1$), we can
estimate the evolution of both $\dNa$ and $\etaL$ by demanding the
vanishing of the \rhs of~\refs{eq:BEN,eq:BEL}. Although such an
estimation can only be done numerically, it can still be helpful as it
shows that the initial conditions do not change the lepton asymmetry
at low temperatures.

\begin{figure}[ht!]
\centering
	\begin{subfigure}[]{0.48\textwidth}
		\includegraphics[width=1.\textwidth]{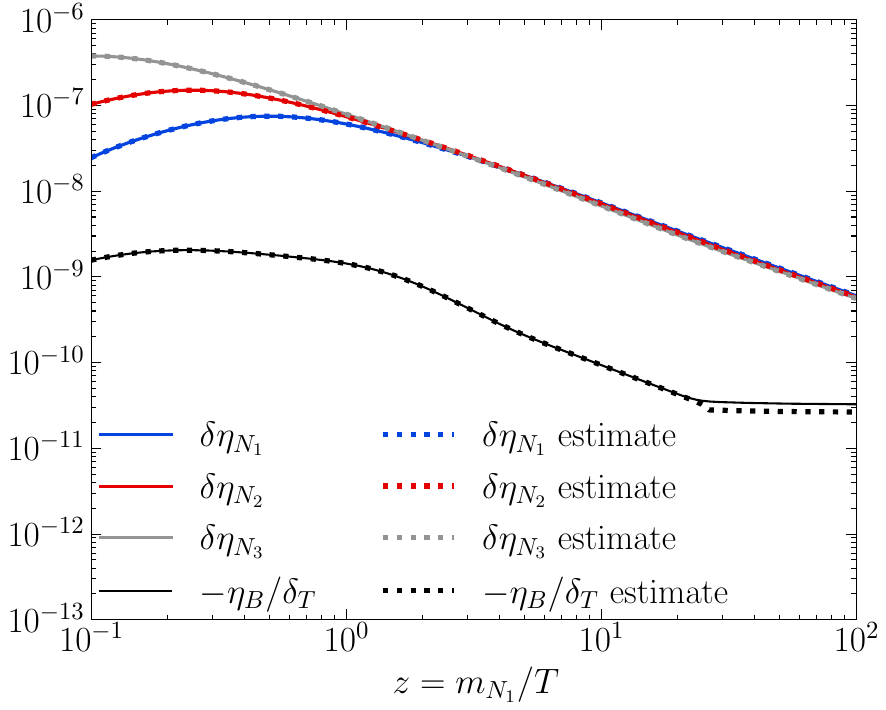}
		\caption{}
		\label{fig:num_vs_approx-no_degen}
	\end{subfigure}
	\begin{subfigure}[]{0.48\textwidth}
		\includegraphics[width=1.\textwidth]{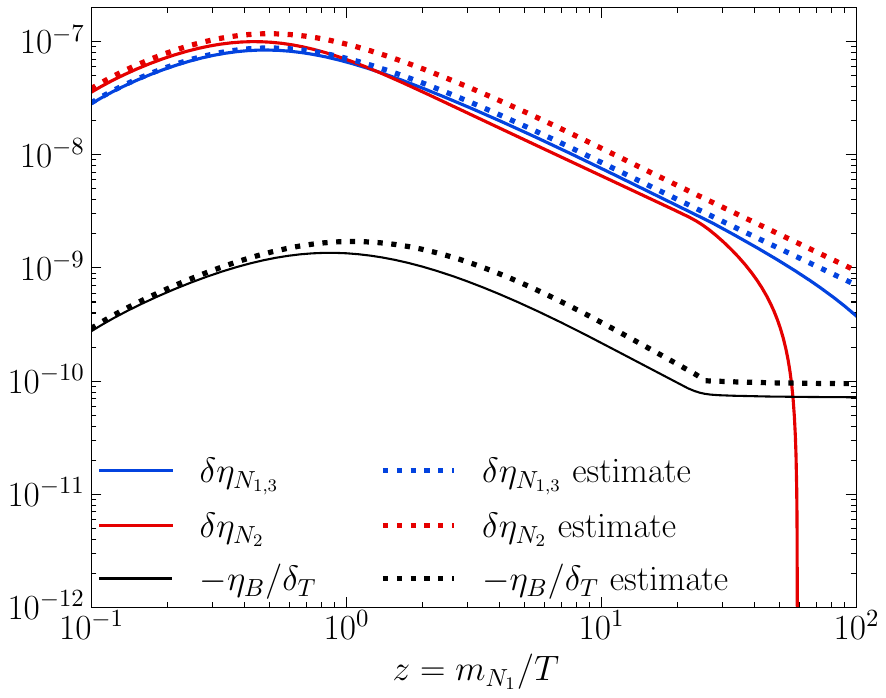}
		\caption{}
		\label{fig:num_vs_approx-degen}
	\end{subfigure}
    \caption{Numerical and approximate solutions of the
      BEs~(\ref{eq:BEN}) and~(\ref{eq:BEL}) for two points in our
      parameter space with $m_{N_1}=m_{N_2}/2=m_{N_3}/4 = 700~\GeV$
      (left) and $m_{N_3} \approx m_{N_2}\approx m_{N_1} = 700~\GeV$
      (right), and $|\h^{\nu}_{ij}| \approx 3 \times 10^{-4}$. The
      lines correspond to the numerical solutions for $\delta_{N_{1}}$
      (blue), $\delta_{N_{2}}$ (red), $\delta_{N_{3}}$ (grey), and
      $\etaL/\delta_1$ (black), while the dotted lines of the same
      colour show the corresponding estimate.} 
    \label{fig:num_vs_approx}
\end{figure}
In \Figs{fig:num_vs_approx-no_degen}, we show the solution of the
BEs~\refs{eq:BEN,eq:BEL} for $m_{N_1} = 700~\GeV$, $m_{N_2} =
1.4~\TeV$, $m_{N_3} = 2.8~\TeV$, and  $|\h^{\nu}_{ij}| \approx 3
\times 10^{-4}$. We note that this is away from the resonant region,
so $\delta_T \ll 1$. As this results in a suppressed back-reaction
term of~\refs{eq:BEN}, the numerical estimates turn out to agree with
the numerical solution of the BEs, even at high temperatures. This
suggests that the addition of the $2 \to 2$ processes pushes the
system towards its attractor solution faster. Finally, as can be seen,
the freeze-out is not identified correctly, and the resulting baryon
asymmetry is slightly underestimated. 

For scenarios with large $\delta_\alpha$, the approximation deviates
from the attractor solution, as the BEs~\refs{eq:BEN,eq:BEL} are no longer
independent due to the lepton back-reaction contribution
to~\refs{eq:BEN}. In \Figs{fig:num_vs_approx-degen} we show the
evolution of $\dNa$ and $\eta_B$ for $m_{N_1} = 700~\GeV$ (with
$m_{N_{2,3}} $ satisfying the tri-resonant
condition~\refs{eq:resonant_condition}) and $|\h^{\nu}_{ij}| \approx 3
\times 10^{-4}$, which results in $\delta_T \sim -1$ (Benchmark
$\mathbf{B}$ in Appendix \ref{app:deltah_values}). 
In turn, the back-reaction term in~\refs{eq:BEN} gives a non-trivial
contribution to the evolution of $\dNa$.  Consequently, for $z \gg 1$,
$\delta\eta_{N_{1,2,3}}$ decrease at a higher rate, resulting in a
discrepancy between the numerical and approximate solutions.      
Note that, as shown in \Figs{fig:cp_asymmetry},  we have $\delta_1 \approx
\delta_3$, while $\delta_2$ yields the dominant contribution to the CP
asymmetry. This is reflected in \Figs{fig:num_vs_approx-degen}, as
$\delta \eta _{N_2}$ begins to fall at lower $z$ than $\delta \eta
_{N_{1,3}}$.

\subsubsection{The Effect of Varying Relativistic Degrees of Freedom}
\begin{figure}[t!]
\centering
	\includegraphics[width=0.6\textwidth]{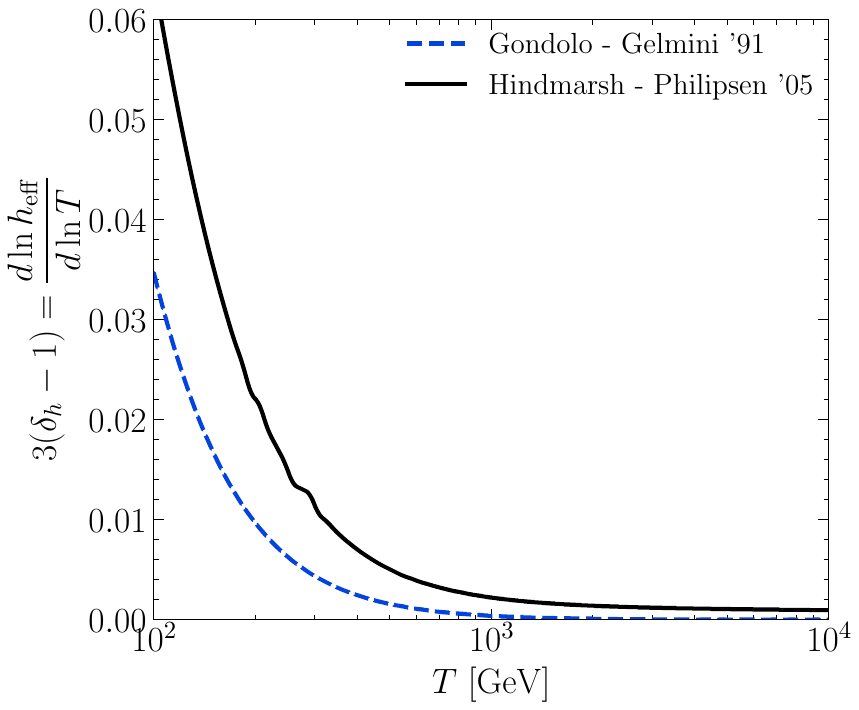}
	\caption{The derivative of $\ln\heff$ with respect to $\ln T$ for $100~\GeV \leq T\leq 10~\TeV$. The black (solid) line corresponds to the tabulated values of $\heff$ given in~\cite{Hindmarsh:2005ix}, while the blue (dashed) line corresponds to~\cite{Gondolo:1990dk}.}
	\label{fig:dlogheffdlogT}
\end{figure}

As already mentioned, we have taken into account the temperature
dependence of the effective relativistic dofs of the plasma, which
introduces a dependence on $d\ln\heff/d\ln T$ in
both~\refs{eq:BEN,eq:BEL}. 
In \Figs{fig:dlogheffdlogT} we show $d\ln\heff/d\ln T$ as a function
of the temperature in the range $100~\GeV \leq T\leq 10~\TeV$. The two
lines correspond to the tabulated values given
in~\cite{Hindmarsh:2005ix} (in black) and~\cite{Gondolo:1990dk} (in
blue). 
Despite the seemingly small deviation from zero, the effect of a
non-vanishing derivative of $\heff$ is, in general, not negligible. In
particular, for $\deltah > 1$ the last term of~\refs{eq:BEN} can
dominate, which can result in a negative $\dNa$. If this happens at
temperatures close to  $\Tsph$, $\dNa$ does not have time to ``bounce"
to positive values. Then, since $\etaL$  depends on $\dNa$, $\eta_B$
can also obtain a negative value at $T = \Tsph$. 
\begin{figure}[t!]
\centering
	\begin{subfigure}[]{0.48\textwidth}
		\includegraphics[width=1.\textwidth]{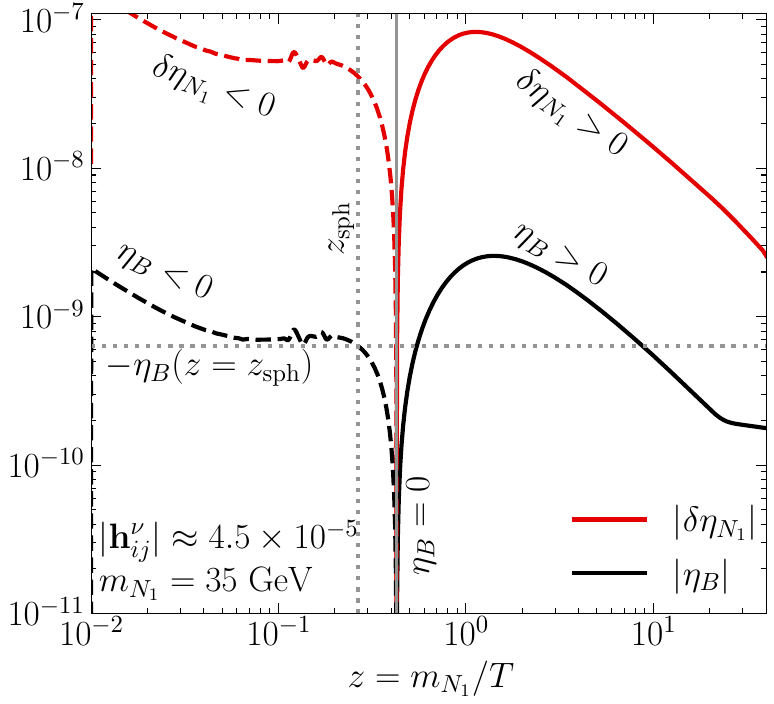}
		\caption{}
		\label{fig:etaB_negative}
	\end{subfigure}
	\begin{subfigure}[]{0.48\textwidth}
		\includegraphics[width=1.\textwidth]{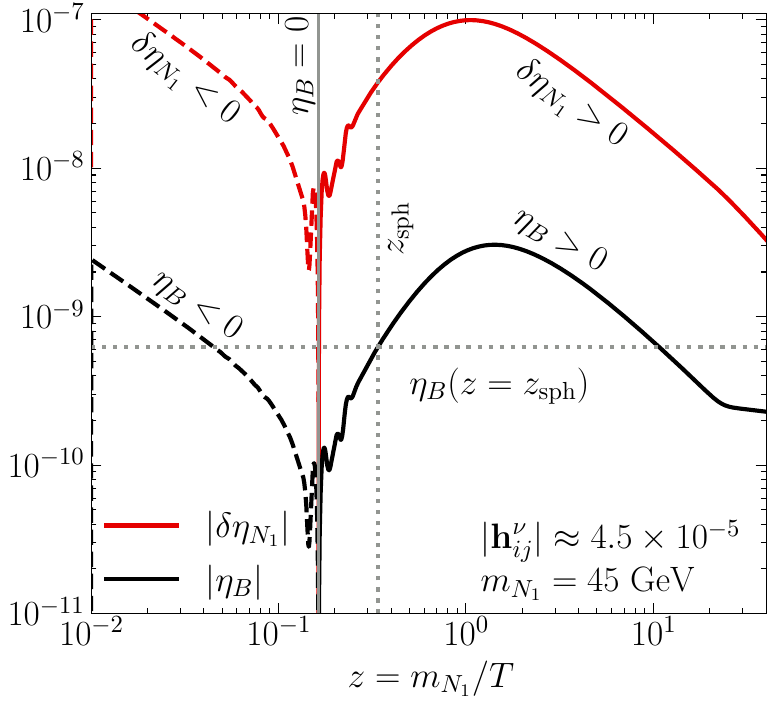}
		\caption{}
		\label{fig:etaB_positive}
	\end{subfigure}
	\centering
    \caption{Evolution of $\delta_{N_1}$ (red) and $\etaL$ (black) for
      $m_{N_1} = 35~\GeV$ (a) and $m_{N_1} = 45~\GeV$ (b), and
      $|\h^{\nu}_{ij}| \approx 4.5 \times 10^{-5}$. The black and red
      solid (dashed) lines show where $\eta_B$ and $\delta_{N_1}$are
      positive (negative), the solid grey lines shows the points where
      $\eta_B = 0$. The vertical dotted grey lines correspond to
      $\zsph=m_{N_1}/\Tsph$, while the horizontal ones show the value
      of $|\eta_B|$ at $z=\zsph$.} 
    \label{fig:etaB_sign}
\end{figure}
This behaviour is observed in \Figs{fig:etaB_sign}, where we show the
evolution of $\delta_{N_1}$ (in red) and $\etaL$ (in black) for
$m_{N_1} = 35~\GeV$ (\Figs{fig:etaB_negative}) and $m_{N_1} = 45~\GeV$
(\Figs{fig:etaB_positive}). In both
\Figs{fig:etaB_negative,fig:etaB_positive}, the solid (dashed) lines
show the regime where the corresponding quantities are positive
(negative), and the vertical solid grey lines show the points where
$\eta_B = 0$. The vertical dashed grey line corresponds to $T=\Tsph$,
while the horizontal one displays the value of $\eta_B$ at $z=\zsph$.    
We observe that, although the curves in both figures are similar, the
prediction for the baryon asymmetry is considerably
different. Particularly, in both
\Figs{fig:etaB_negative,fig:etaB_positive}, the quantities are
negative for $z \ll 1$ and they change sign close to $z \sim
10^{-1}$. For $m_{N_1}=35~\GeV$, this sign change would occur after
the sphalerons decouple. Therefore the generated BAU is
negative. Conversely, for $m_{N_1}=45~\GeV$, the sphaleron freeze-out
occurs after $\eta_B$ becomes positive, and thus the generated BAU is
positive. Moreover, we should note that this behaviour is stable under
perturbing the initial conditions of both $\dNa$ and $\etaL$, as the
system reaches quickly its attractor solution.  

Therefore, there seems to be a mass scale, below which the resulting
$\eta_B$ is negative, which also depends on the values of $\heff$ and
its derivative. 
\begin{figure}[t!]
\centering
	\begin{subfigure}[]{0.48\textwidth}
		\includegraphics[width=1.\textwidth]{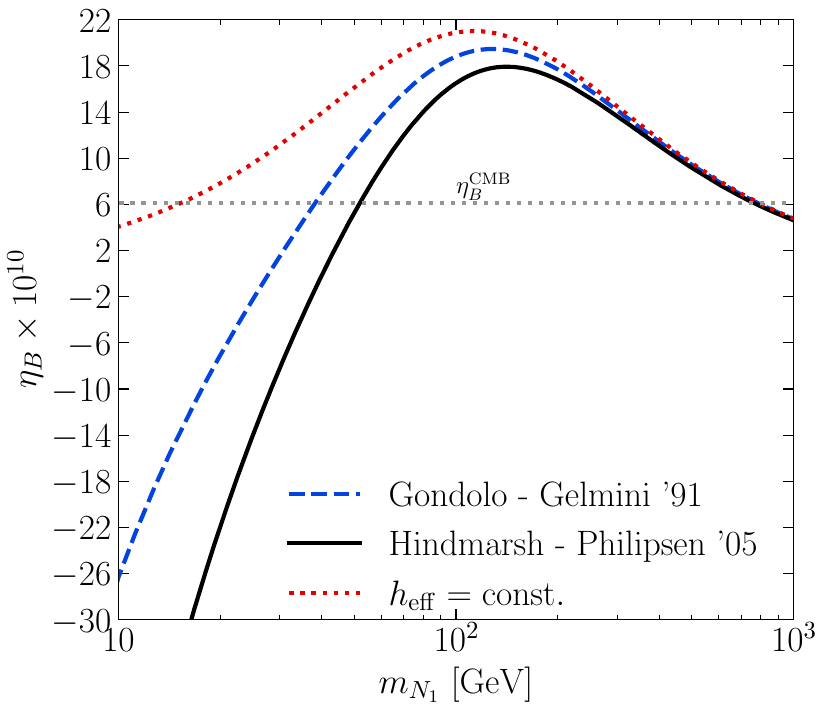}
		\caption{}
		\label{fig:heff_effect}
	\end{subfigure}
	\begin{subfigure}[]{0.48\textwidth}
		\includegraphics[width=1.\textwidth]{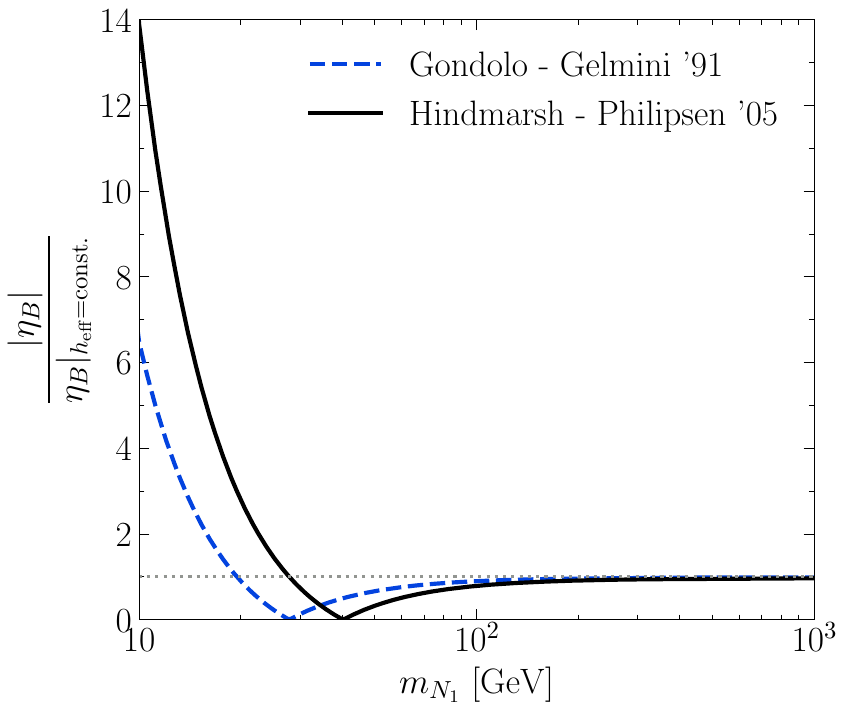}
		\caption{}
		\label{fig:heff_ratios}
	\end{subfigure}
	\centering
    \caption{\textit{Left panel:} The resulting $\eta_B$ for
      $|\h^{\nu}_{ij}| \approx 3 \times 10^{-4}$ in the tri-resonant
      scenario  as a function of $m_{N_1}$ for $\heff$ as given
      in~\cite{Hindmarsh:2005ix} (black),~\cite{Gondolo:1990dk}
      (blue), and taking $\heff = {\rm const.} \approx 105$ (red). The
      grey dotted line shows $\eta_B^{\rm CMB} = 6.104 \times
      10^{-10}$. \textit{Right panel:} The ratio of $|\eta_B|$ with
      varying $\heff$, the black (blue) line corresponds
      to~\cite{Hindmarsh:2005ix} (\hspace{1sp}\cite{Gondolo:1990dk}),
      with respect to $\heff = {\rm const.}$ The grey dotted line
      indicates $|\eta_B|/\eta_B|_{\heff = {\rm const.} }=1$. } 
    \label{fig:heff_var}
\end{figure}
In \Figs{fig:heff_effect} we show the dependence of $\eta_B$ for
$|\h^{\nu}_{ij}| \approx 3 \times 10^{-4}$ in the tri-resonant
scenario using three different forms of $\heff$: (i)~using data given
in~\cite{Hindmarsh:2005ix} (in black), (ii)~the tabulated $\heff$
provided in~\cite{Gondolo:1990dk} (in blue), and (iii)~taking $\heff =
{\rm const.} \approx 105$ (in red). We observe that $\eta_B$ depends
heavily on the derivative of $\heff$ for $m_{N_1} \lesssim
100~\GeV$. In particular, the commonly used assumption, $\heff = {\rm const.}$,
results in overall larger baryon asymmetry today, while the cases with
varying $\heff$ are lower. In fact, larger values of $d\ln\heff/d\ln
T$ imply a smaller $\eta_B$. Moreover, at around $m_{N_1} = 40~\GeV$,
$\eta_B$ becomes negative, which means that this is a scale below
which the CP asymmetries need to change their sign. A positive
$\eta_B$ can be obtained by changing $\mathbf{h}_+^\nu$ to its CP
conjugate, \eg by $\omega \to \omega^*$.  We should stress that the
regime $m_{N_1} \lesssim 100~\GeV$ needs to be carefully studied along
with the values of $\heff$ at temperatures larger than $\Tsph$, as the
baryon asymmetry is very sensitive on the derivative of $\heff$. 
This sensitivity can also be seen in \Figs{fig:heff_ratios}, where we
show the ratio between $|\eta_B|$ with varying $\heff$ (again, black
corresponds to~\cite{Hindmarsh:2005ix} and blue
to~\cite{Gondolo:1990dk}) with respect to $\heff = {\rm const}$. In
this figure, we observe again that $m_{N_1} \lesssim 100~\GeV$ is
sensitive to the derivative of $\heff$, as both lines deviate
considerably from $1$ as well as from each other by a factor of $\sim
2$ for $m_{N_1} \lesssim 40~\GeV$. In the regime $m_{N_1} \lesssim 40~\GeV$, the baryon asymmetry would receive contributions from other effects, \eg,
from coherent heavy neutrino
oscillations~\cite{Akhmedov:1998qx,Asaka:2005pn,Asaka:2011wq,
  Shuve:2014zua,Hambye:2016sby,Drewes:2016gmt,BhupalDev:2014pfm,Klaric:2021cpi}.
  The impact of the derivative of $\heff$ on the dynamics of the
  baryon asymmetry would not be affected by the inclusion of these
  effects, as they would introduce another source of CP asymmetry at most comparable to the one considered in the BE (\ref{eq:BEL}). As
  a consequence, previous analyses that do not take the effect of $d\ln\heff/d\ln
T$ into account must be revisited accordingly.

\section{Results}\label{sec:results}
\setcounter{equation}{0}

\begin{figure}[t!]
	\begin{subfigure}[]{0.48\textwidth}
		\includegraphics[width=1.\textwidth]{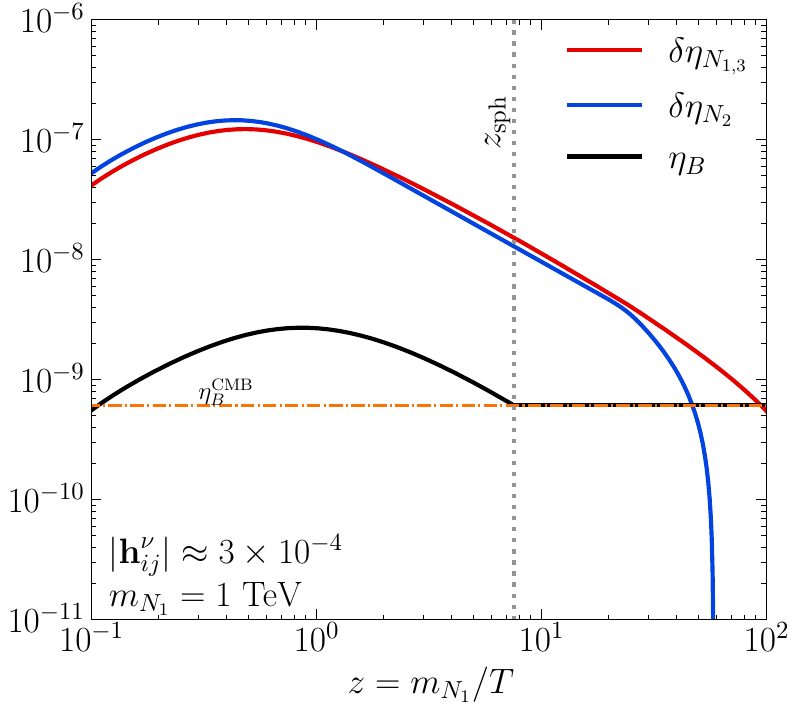}
		\caption{}
		\label{fig:evolution_high}
	\end{subfigure}
	\begin{subfigure}[]{0.48\textwidth}
		\includegraphics[width=1.\textwidth]{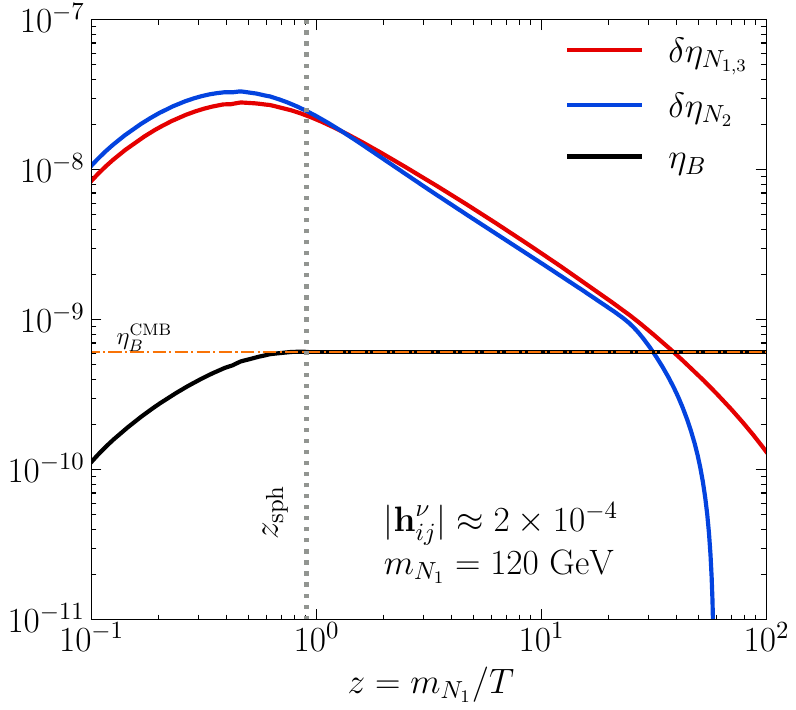}
		\caption{}
		\label{fig:evolution_low}
	\end{subfigure}
	\centering
    \caption{Evolution plots for the baryon asymmetry $\eta_B$ (black
      solid line) and the deviation from equilibrium of the neutrino
      densities $\delta\eta_{N_\alpha}$ (blue and red solid
      lines). The choice for the values of the mass of the lightest
      singlet neutrino and the scale of the Yukawa coupling can be
      seen on each panel, and the grey (dotted) line indicates the
      value $z=\zsph$ at which the sphaleron processes freeze out. The
      orange dot-dashed line indicates the observed value of the
      baryon asymmetry of $\eta^{\rm CMB}_B = 6.104\times 10^{-10}$.} 
    \label{fig:evol_plots}
\end{figure}

We present the numerical solutions to the BEs as shown
in~\refs{eq:BEN,eq:BEL} for the benchmark model defined by the
rescaled Yukawa matrix presented in~\refs{eq:rescaledyukawa}, and for
a tri-resonant singlet neutrino spectrum. In the following analysis,
we restrict ourselves to heavy neutrino masses above $40~\GeV$. Below
this mass scale, our approach is more limited due to the fact that by
ignoring thermal masses, we do not account for phase space suppression
effects and their impact on the leptonic asymmetries. Besides these
thermal effects, a more detailed
treatment~\cite{BhupalDev:2014pfm,BhupalDev:2014oar,Kartavtsev:2015vto,Racker:2020avp,Jukkala:2021sku,Racker:2021kme}
would require us to incorporate additional CP violating effects
induced by the coherent oscillation of heavy
neutrinos~\cite{Akhmedov:1998qx,Asaka:2005pn,Asaka:2011wq,
  Shuve:2014zua,Drewes:2016gmt,Klaric:2021cpi}, along with those
effects that come from their CP-violating
decays~\cite{Pilaftsis:1997jf,Pilaftsis:1997dr,Pilaftsis:2003gt}. Finally, we note that
for low singlet neutrino masses, the necessary\- CP violation could
originate from Higgs decays into a singlet neutrino and a lepton
doublet when the thermal effects of the plasma are
considered~\cite{Hambye:2016sby,PhysRevD.96.015031}.

As discussed in the previous section, the scattering terms generate a
delay in the onset of the maximum of the baryon asymmetry, which
modifies the shape of the curve for the baryon asymmetry evolution
depending on the mass of the singlet
neutrinos. Figure~\ref{fig:evolution_high} shows the evolution of the
baryon asymmetry for $|\h^\nu_{ij}|=3\times10^{-4}$ and
$m_{N_1}=1~\TeV$ (labeled as Benchmark {\bf C} in
Appendix~\ref{app:deltah_values}), and it demonstrates that for $\TeV$
singlet neutrinos, the baryon asymmetry is reached by the freeze-out
of the lepton asymmetry after the maximum value is
reached. Figure~\ref{fig:evolution_low} presents the evolution for
$|\h^\nu_{ij}|=2\times10^{-4}$ and $m_{N_1}=120~\GeV$ (Benchmark {\bf
  D} in Appendix \ref{app:deltah_values}), and illustrates how the
generation of the baryon asymmetry happens at the maximum when the
mass of the heavy neutrinos becomes lower. In both panels, we selected
the initial conditions $\eta_L(z_0) = 0$ and $\delta\eta_{N_\alpha}
(z_0)=0$ for $\alpha=1,2,3$, with $z_0=10^{-2}$. However, due to the
attractive nature of the solution to the BEs, the evolution of the
baryon asymmetry remain effectively unchanged for any other reasonable
choice. Also, since $\delta_2$ is the largest CP asymmetry,
$\delta\eta_{N_{2}}$ deviates significantly from
$\delta\eta_{N_{1,3}}$ at high values of $z$, as outlined in
Section~\ref{sec:approx}.  
\begin{figure}[t!]
\centering
\begin{subfigure}[]{0.49\textwidth}
	\includegraphics[width=1.0\textwidth]{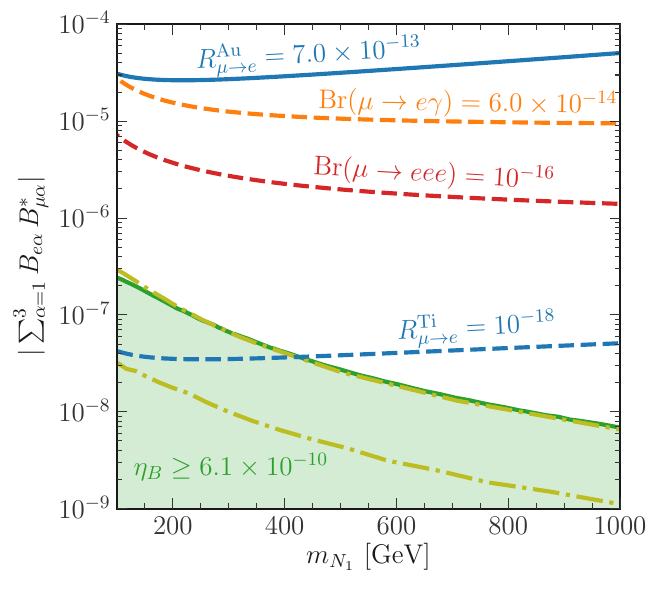}
	\caption{}
	\label{fig:lfv_lims}
\end{subfigure}
\begin{subfigure}[]{0.49\textwidth}
	\includegraphics[width=1.0\textwidth]{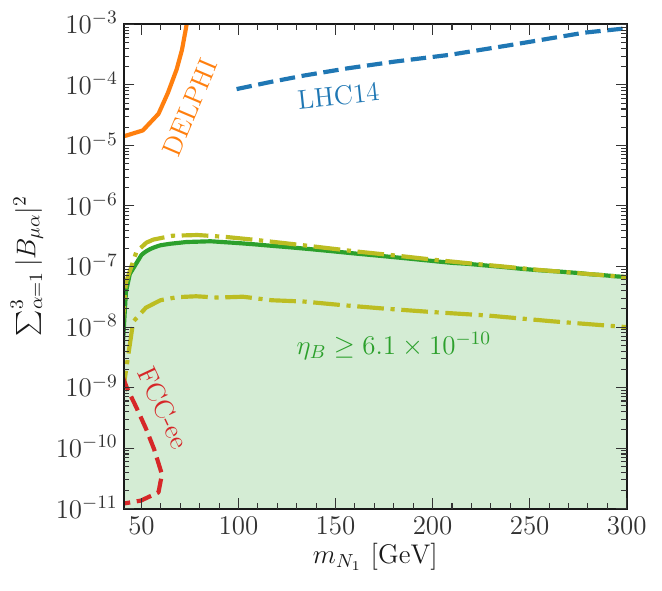}
	\caption{}
	\label{fig:collider_lims}
\end{subfigure}
\centering
\caption{Parameter space for the TRL model, including current limits
  (solid lines) and projected sensitivities of future experiments
  (dashed lines). \textit{Left panel:} Projected sensitivities of cLFV
  searches for $\mu\rightarrow e\gamma$ (orange dashed line),
  $\mu\rightarrow eee$ (dashed red line), coherent $\mu\rightarrow e$
  conversion in titanium (dashed blue line), and current limits from
  searches for coherent $\mu\rightarrow e$ conversion in gold (solid
  blue line). These curves are based on the limits presented in
  (\ref{eq:current_constraints}) and
  (\ref{eq:projected_sensitivities}). \textit{Right panel:} Projected
  sensitivities for collider searches at LHC$14$ (blue dashed line),
  FCC-ee (red dashed line), and current limits from DELPHI (orange
  solid line). For details, see text. In both panels, the green region
  indicates points in the parameter space where leptogenesis can lead
  to the observed value for the baryon asymmetry, where the green
  solid line corresponds to the points that reproduce exactly the
  observed value for a tri-resonant model. 
  The upper and lower yellow dot-dashed lines were obtained by scaling the total CP asymmetry $\delta_T$ by a factor of 2 and 0.1 respectively, and matching the observed baryon asymmetry. They represent an estimate of the uncertainty on the calculation of the solid green line due to oscillations.}
\label{fig:parameter_space}
\end{figure}

Figure~\ref{fig:parameter_space} shows the parameter space for the TRL
model on the $\sum_\alpha B_{l\alpha}B^*_{k\alpha}$ vs. $m_{N_1}$
plane. For this figure, we set $a=b=c$ in (\ref{eq:z6yukawa}). Additionally, we assume that the masses are in consecutive resonance, as defined in (\ref{eq:resonant_condition}). For definiteness, we take as initial conditions $\eta_L(z_0) = 0$ and $\delta\eta_{N_\alpha}
(z_0)=0$, with $z_0=10^{-2}$, although our results are largely independent of the initial conditions. 
The region of parameter space that leads to a successful
generation of the baryon asymmetry is depicted in green, where the
solid green line indicates the value for which the obtained baryon
asymmetry is equal to the observed value of $\eta^{\rm CMB}_B =
6.104\times 10^{-10}$. The green dashed line indicates an estimate of this curve when an additional source of CP violation, generated by heavy neutrino oscillations, is included. 
For a fixed mass, the green region below the
line yields, in principle, a higher value for $\eta_B$, but a
relaxation of the tri-resonant condition (\ie~an increase or decrease
in mass differences) can adjust the precise observed value.

In Figure \ref{fig:parameter_space}, we have included an upper and a lower yellow dot-dashed line, obtained by scaling the total CP asymmetry $\delta_T$ by a factor of 2 and 0.1, respectively. They represent an estimate of the theoretical uncertainties associated to the heavy neutrino oscillation effects that were ignored in the BEs (\ref{eq:BEN}) and (\ref{eq:BEL}). This estimate was made by assuming, in line with \cite{BhupalDev:2014pfm}, that the contribution from oscillations denoted there by $\delta\eta_{L,{\rm osc}}$ acts as a source of CP asymmetry which is distinct from and additive to the one that originates from mixing, $\delta\eta_{L,{\rm mix}}$, which we only consider here. It is necessary to point out, however, that a full three neutrino oscillation formula will differ from the approximate expression for $\delta\eta_{L,{\rm osc}}$ as presented in \cite{BhupalDev:2014pfm} [c.f.~(5.21) therein], potentially allowing for constructive or destructive contributions to CP violating effects in the tri-resonant regime. This different treatment reflects the lack of consensus in the literature concerning whether the mixing of heavy neutrinos is contained within the oscillation phenomenon (e.g. \cite{Garbrecht:2011aw}), or are two different mechanisms
\cite{BhupalDev:2014pfm,BhupalDev:2014oar}. Likewise, mixing and oscillation effects could potentially interfere, a conclusion that is supported by the results presented in \cite{Kartavtsev:2015vto}. On the basis of this ongoing debate, we relegate the study of tri-resonant heavy neutrino oscillation effects to a future work.

The left panel (\Figs{fig:lfv_lims}) shows the sensitivity estimates
and limits on Majorana neutrino models from searches for cLFV
transitions involving muons. We only include the lines that lie close
to the parameter space that leads to sufficient baryogenesis, and
ignore current limits besides the one from coherent $\mu\rightarrow e$
conversion in gold. As it can be seen, the green area is currently far
from  the region of cLFV detection, and the only experiment that could
probe this parameter space in the future is PRISM by searching for
coherent $\mu\rightarrow e$ conversion in titanium.  
The right panel (\Figs{fig:collider_lims}) shows projected and current
limits from collider observables. The estimate denoted by LHC$14$
(blue dashed line) presents conservative projections for the
sensitivity to the process $pp\rightarrow N\ell^\pm jj$ at the LHC
with $300~{\rm fb}^{-1}$ data operating at $\sqrt{s} =
14~\TeV$~\cite{Deppisch:2015qwa,Dev:2013wba}. The DELPHI line (orange,
solid) represents 95\% C.L. limits found by comparing LEP data with
the prediction for signals of decaying heavy neutrinos that are
produced via $Z\rightarrow N\nu_L$~\cite{DELPHI:1996qcc}. Similar
limits have been derived by the L3
collaboration~\cite{L3:1992xaz}. The red dashed line shows the
sensitivity to the same signals at the Future Circular Collider (FCC)
for electron-positron collisions assuming normal order of the light
neutrino spectrum, and considering the lifetime of the heavy
neutrinos~\cite{Blondel:2014bra}. 

In summary, Figure~\ref{fig:lfv_lims} highlights the potential of
PRISM to probe the parameter space of our leptogenesis model in the
mass range below $400~\GeV$. On the collider front,
Figure~\ref{fig:collider_lims} shows that high luminosity
$Z$-factories could probe the parameter space in a narrow range of
masses, but for remarkably low values of the light-to-heavy neutrino
mixings. Below the mass range we analyse, an extensive portion of the
parameter space is ruled out by searches for heavy neutrinos that are
produced in fixed target experimental facilities
(e.g~\cite{PIENU:2017wbj,PIENU:2019usb,NA62:2020mcv,NA62:2021bji,Bernardi:1985ny,Bernardi:1987ek,NuTeV:1999kej,T2K:2019jwa,MicroBooNE:2019izn,NOMAD:2001eyx})
or in atmospheric
showers~\cite{Kusenko:2004qc,Coloma:2019htx,Arguelles:2019ziu}, while
future upgrades promise a significant gain in sensitivity for the
heavy neutrino parameter space. This further motivates a complete
analysis including heavy neutrino oscillation effects, which could
reveal a viable parameter space within the reach of these
experiments. Above the $Z$ pole, the LHC $14$ projection lies far
above the region where leptogenesis is successful. In models with two
singlet neutrinos, an analysis searching for LNV lepton-trijet and
dilepton-dijet signatures in future electron-positron, proton-proton,
or electron-proton colliders shows that a sensitivity close to
$10^{-6}$ could be achieved in the range of a few hundred $\GeV$
\cite{Antusch:2016ejd}. However, this is still an order of magnitude
above our prediction for the viable leptogenesis parameter space, and
a potential improvement on the sensitivity by the addition of a third
singlet neutrino would require a dedicated analysis. A recent
extensive review of current bounds and projections, including several
exclusion lines that we omit in the presentation of our results, can
be found in~\cite{Abdullahi:2022jlv}. For bounds and projections on
multi-TeV heavy neutrinos, the interested reader may consult the
recent results communicated in \cite{Calderon:2022alb}.

\section{Conclusions}\label{sec:conclusions}
\setcounter{equation}{0}

We have studied a class of leptogenesis models where the smallness
of the light-neutrino masses is accounted for by approximate discrete
symmetries, such as $\mathbb{Z}_3$ or $\mathbb{Z}_6$ symmetries. The new feature of this
class of models is that they may naturally give rise to three nearly
degenerate heavy Majorana neutrinos that can strongly mix with one
another and have mass differences comparable to their decay widths. In
particular, we have shown how such tri-resonant heavy neutrino systems
can lead to leptonic CP asymmetries that are further enhanced than
those obtained in the frequently considered bi-resonant approximation.
In this context, this enhanced mechanism of leptogenesis was termed
Tri-Resonant Leptogenesis. 

Following~\cite{Pilaftsis:2005rv}, we have formulated the BEs for TRL
by considering chemical potential corrections, as well as by keeping
the temperature dependence of the effective relativistic dofs of the plasma
($\heff$ and $\geff$). We have found that the latter may result in significant
corrections to the heavy neutrino number density and lepton asymmetry
BEs.  To the best of our knowledge, these corrections have not been
taken into account before in the numerical estimates of the
baryon-to-photon ratio $\eta_B$ in thermal leptogenesis.

After performing a careful numerical study of the solutions to the
evolution equations, we have explicitly demonstrated that for
$m_N\stackrel{<}{{}_\sim} 100 ~\GeV$, the effect of the derivative of
$\heff (T)$ with respect to the temperature $T$ of the plasma has an
important influence on the evolution of the baryon asymmetry
$\eta_B$ in the Universe. Moreover, an accurate determination of
$\heff (T)$ will reduce the uncertainty in the predictions for the
BAU.
In addition, as illustrated in Figure~\ref{fig:parameter_space}, our approach to the BEs may be limited due to the uncertainties pertaining the omission of heavy neutrino oscillations, and a complete treatment would need to account for these phenomena. Given the alternate approaches to the treatment of neutrino oscillation and mixing effects, we have decided to postpone such considerations for later work.


In the TRL models that we have been studying here, the allowed
parameter space that leads to successful leptogenesis gets
significantly enlarged as compared to the expectation from ordinary
seesaw models. Furthermore, a part of this parameter space will be
probed by several projected experiments that include both cLFV and
collider observables. Extensions to this model, via its
supersymmetrisation, could lead to a considerable expansion of the
leptogenesis parameter space due to the potential occurrence of
additional cancellations that allow higher values of light-to-heavy
neutrino mixings~\cite{CandiadaSilva:2020hxj}. Likewise, the possible
existence of more than three nearly degenerate heavy neutrinos can
trigger a much more involved multi-resonant dynamics. Hence, requiring
successful multi-Resonant Leptogenesis may imply a further relaxation
of the stringent constraints on the theoretical parameters of such
models.  Finally, the inclusion of flavour effects may enhance the
prospects of observable cLFV and LNV in future experiments.  We aim
to return and study some of the issues mentioned above in the near
future.

\vspace{-3mm}
\subsection*{Acknowledgements} 
\vspace{-3mm}

\noindent
The work of AP and DK is supported
in part by the Lancaster-Manchester-Sheffield Consortium for
Fundamental Physics, under STFC Research Grant ST/T001038/1.
The work of PCdS is funded by Agencia Nacional de Investigaci\'on y
Desarrollo (ANID) through the Becas Chile Scholarship
No.~72190359. TM acknowledges support from the STFC Doctoral Training
Partnership under STFC training grant ST/V506898/1.

\newpage
\setcounter{section}{0}
\section*{Appendix}
\appendix

\renewcommand{\theequation}{\Alph{section}.\arabic{equation}}
\setcounter{equation}{0}  

\section{Impact of {\boldmath $T$}-dependent {\boldmath $h_{\rm eff}$} on BEs}
\label{app:BE_dh}
\setcounter{equation}{0}

In order to derive~\refs{eq:BEN,eq:BEL}, we begin by writing down the general
form of a BE in an isotropically expanding FRW Universe,
\begin{equation}
    \dfrac{d n}{d t}\: +\: 3\,H\,n\ =\ {\cal C} \;,
    \label{eq:BE_Gen}
\end{equation}
with ${\cal C}$ representing the relevant collision terms. In order to be able to
solve such an equation, we rewrite it in terms of the temperature $T$,
instead of the cosmic time $t$. This can be done by assuming conservation of the
comoving entropy $(S=sa^3)$, which implies $d s/ d t = - 3 \,H \,
s$.\footnote{In principle, we 
  could perform the variable transformation using $d\rho / dt = -3
  H(\rho + p) = -3  H s T  $. However, this approach produces the same
  result in a less transparent manner.} The~latter observation enables us to perform the
change of variables 
\begin{equation}
    \dfrac{d }{d t} = \dfrac{d s}{d t}\dfrac{d T}{d s} \dfrac{d }{d T}
    = H \, \deltah^{-1}  \dfrac{d }{d \ln z }\;, 
    \label{eq:t_to_T}
\end{equation}
where $\delta_h$ was defined in~\eqref{eq:deltah_def}, $z= M/T$, and $M$
is some convenient mass scale, which 
in~\refs{eq:BEN,eq:BEL} is chosen to be~$\mNa$. 

For the BE that describes the evolution of the lepton
asymmetry~\refs{eq:BEL}, we write the \lhs of~\refs{eq:BE_Gen} as 
\begin{equation}
    \dfrac{d n_L}{d t} +3\,H\,n_L\, =\, 
    \etaL \, \dfrac {n_{\gamma}}{d t} +\dfrac{\eta_L}{d t} \, n_{\gamma} +3\,H\,\etaL\,n_\gamma\,
    =\, n_{\gamma} \, H \, \deltah^{-1} \bigg[  \dfrac{d \etaL}{d \ln z}
    +3 \etaL \Big(\deltah-1\Big) \bigg]\;,
    \label{eq:BEL_lhs}
\end{equation}
where we have used~\refs{eq:t_to_T} and
$d n_\gamma / d t= -3\,H\,\deltah^{-1} \,n_\gamma$.  Identifying
${\cal C}$ with the
collision terms given in~\cite{Pilaftsis:2005rv}, we obtain the
evolution equation for the lepton asymmetry~\refs{eq:BEL}.

Following the same steps as above, the heavy neutrino BE~\refs{eq:BEN} becomes
\begin{equation}
    \dfrac{d n_{N_\alpha}}{d t} +3\,H\, n_{N_\alpha}\, =\, 
     n_{\gamma} \, H \, \deltah^{-1} \bigg[  \dfrac{d \etaNa}{d \ln z}
     +3 \etaNa \Big(\deltah-1\Big) \bigg]\;.
    \label{eq:BEN_lhs_init}
\end{equation}
We can then express~\refs{eq:BEN_lhs_init} in terms of $\dNa$ as 
\begin{equation}
    \dfrac{d n_{N_{\alpha}}}{d t} +3\,H\,n_{N_{\alpha}}\, =\,  
    H \, \deltah^{-1} \, n_\gamma \, \eta_{N_{\alpha}}^{\rm eq}  
    \bigg[\dfrac{d \dNa }{d \ln z }
    +\Big(\dNa +1\Big)\dfrac{d \ln \eta_{N_{\alpha}}^{\rm eq} }{d \ln z }  
    +3\Big(\dNa+1\Big)\Big(\deltah-1\Big)\bigg] \;,
    \label{eq:BE_dY_lhs}
\end{equation}
where 
\begin{align}
    &\dfrac{d \ln \eta_{N_{\alpha}}^{\rm eq}}{d \ln z }\,=\, - \dfrac{\mNa}{T} \dfrac{K_1(\za)}{K_2(\za)} \;.
    \label{eq:Yeq}
\end{align}
Once again, substituting ${\cal C}$ for the collision terms
evaluated in~\cite{Pilaftsis:2005rv}, we arrive at~\refs{eq:BEN}.

\section{Form Factors for cLFV Processes}
\label{app:formfactors}
\setcounter{equation}{0}

We list the form factors that appear in the calculation of the cLFV
processes in~\eqref{eq:brmue}, \eqref{eq:brmu3e}
and~\eqref{eq:ratemue}. These depend on the 
light-to-heavy neutrino mixing defined in~\refs{eq:light-heavy_mixing}
and are given by~\cite{Ilakovac:1994kj, Ilakovac:1995km} 
\begin{eqnarray} 
G_\gamma^{\mu e} &=& \sum^3_{\alpha=1} B_{e\alpha}
  B^*_{\mu \alpha}
  G_\gamma(x_{N_\alpha}), \label{eq:Ggamma} \\
  F_\gamma^{\mu e} &=&
  \sum^3_{\alpha=1} B_{e\alpha} B^*_{\mu \alpha}
  F_\gamma(x_{N_\alpha}),
  \label{eq:Fgamma}\\
  F_Z^{\mu e} &=& \sum^3_{\alpha=1} B_{e\alpha}
  B^*_{\mu \alpha}
  \left[F_Z(x_{N_\alpha}) + 2G_Z(x_{N_\alpha},0)\right],
  \label{eq:Fz}\\
  F_{\rm Box}^{\mu euu} &=& \sum^3_{\alpha=1} B_{e\alpha}
  B^*_{\mu \alpha}
  \left[H_{\rm Box}(x_{N_\alpha},0) - H_{\rm Box}(0,0)\right],\label{eq:Fmueuu}\\
  F_{\rm Box}^{\mu edd} &=& -\sum^3_{\alpha=1} B_{e\alpha}
  B^*_{\mu \alpha}
  \left[F_{\rm Box}(x_{N_\alpha},0) -  F_{\rm Box}(0,0)\right],\label{eq:Fmuedd}\\
  F_{\rm Box}^{\mu eee} &=& 2 \: \sum^3_{\alpha=1} B_{e\alpha}
  B^*_{\mu \alpha}
  \left[F_{\rm Box}(x_{N_\alpha},0) - F_{\rm Box}(0,0)\right],
  \label{eq:Fmueee}
\end{eqnarray}
where $x_{N_\alpha}\equiv (m_{N_\alpha}/M_W)^2$.  
In~(\ref{eq:Fz}) and~(\ref{eq:Fmueee}), we neglected terms of order
higher than two in the light-heavy neutrino mixing parameters, since
they do not modify our numerical results, while in~(\ref{eq:Fmuedd})
and~(\ref{eq:Fmueuu}) we ignore the squared modulus of non-diagonal
entries of the CKM matrix.  

The analytic forms of the loop functions that appear in the previous form factors are specified below:
\begin{align}
G_\gamma(x) &= -\frac{x(2x^2+5x-1)}{4(1-x)^3}
  - \frac{3x^3}{2(1-x)^4}\ln x,\\
  F_\gamma(x) &= \frac{x(7x^2-x-12)}{12(1-x)^3}
  \; - \; \frac{x^2(x^2-10x+12)}{6(1-x)^4}\ln x, \\
  F_Z(x) &= -\frac{5x}{2(1-x)}
   - \frac{5x^2}{2(1-x)^2}\ln x,\\
  G_Z(x,0) &= -\frac{x}{2(1-x)}\ln x,\\
  H_{\rm Box}(x,0) &= \frac{4}{1-x} + \frac{4x}{(1-x)^2}\ln x,\\
  F_{\rm Box}(x,0) &= \frac{1}{1-x} + \frac{x}{(1-x)^2}\ln x,
\end{align}
where it is helpful to indicate the limiting values  $H_{\rm  Box}(0,0)=4$  and  $F_{\rm Box}(0,0)=1$.

\section{Benchmark Scenarios}
\label{app:deltah_values}

For each one of the selected benchmarks presented, it is possible to
find numerical solutions\footnote{The perturbations, $\delta
  \mathbf{h}^\nu$, are found by solving~\refs{eq:numassconstraint}
  analytically using {\tt sympy}~\cite{10.7717/peerj-cs.103}.} for the
entries of perturbation matrix $\delta \h^\nu$ such that the model is
in agreement with the observed neutrino oscillation parameters
(see~\refs{eq:neutrino_params}). Here, we present the values of
$\delta \h^\nu$ for four representative points, including the ones
used in the evolution plots shown in Section~\ref{sec:results}. 

\begin{enumerate}[label=\textbf{\Alph*.}]
\item $m_{N_1} = 500 ~\GeV$, $|(\h^\nu_0)_{ij}|=3\times 10^{-3}$,
\begin{equation*}
    \delta \mathbf{h}^\nu = \left( \begin{matrix}
    0 & (6.40 - 8.15\,i)\times 10^{-12} & 0 \\
    (-7.24 + 3.80\,i)\times 10^{-7} & 0 & -(6.91+ 4.37\,i)\times 10^{-7}\\
    -(1.97 + 0.0837\,i)\times 10^{-4} & (0.911- 1.75\,i)\times 10^{-4} & (1.06 - 1.66\,i)\times 10^{-4}
    \end{matrix}\right).
\end{equation*}
\item $m_{N_1} = 700 ~\GeV$, $|(\h^\nu_0)_{ij}|=3\times 10^{-4}$,
\begin{equation*}
    \delta \mathbf{h}^\nu = \left( \begin{matrix}
    0 & (0.898 - 1.14\,i)\times 10^{-10} & 0 \\
    (-8.58 + 4.50\,i)\times 10^{-7} & 0 & -(8.19 + 5.18\,i)\times 10^{-7}\\
    -(2.02 + 0.0578\,i)\times 10^{-5} & (0.910 - 1.75\,i)\times 10^{-5} & (1.01 - 1.69\,i)\times 10^{-5}
    \end{matrix}\right).
\end{equation*}
\item $m_{N_1} = 1 ~\TeV$, $|(\h^\nu_0)_{ij}|=3\times 10^{-4}$,
\begin{equation*}
    \delta \mathbf{h}^\nu = \left( \begin{matrix}
    0 & (1.29- 1.64\,i)\times 10^{-10} & 0 \\
    (-1.03 + 0.539\,i)\times 10^{-6} & 0 & -(9.80 + 6.20\,i)\times 10^{-7}\\
    -(2.03 + 5.19\,i)\times 10^{-5} & (0.910 - 1.75\,i)\times 10^{-5} & (1.00 - 1.70\,i)\times 10^{-5}
    \end{matrix}\right).
\end{equation*}
\item $m_{N_1} = 120 ~\GeV$, $|(\h^\nu_0)_{ij}|=2\times 10^{-4}$,
\begin{equation*}
    \delta \mathbf{h}^\nu = \left( \begin{matrix}
    0 & (2.46-3.14\,i)\times 10^{-11} & 0 \\
    (-3.53 + 1.85\,i)\times 10^{-7} & 0 & -(3.37 + 2.13\,i)\times 10^{-7}\\
    -(1.24 + 0.0421\,i)\times 10^{-5} & -(0.562+ 1.08\,i)\times 10^{-5} & (0.637- 1.04\,i)\times 10^{-5}
    \end{matrix}\right).
\end{equation*}

\end{enumerate}

We note that for all the benchmark models listed above, we have
$m_{N_1}\approx m_{N_2} \approx m_{N_3}$, such that their mass
differences are subleading to the light-neutrino masses
and their mixing.

\newpage
\bibliography{bibs-refs}{}
\bibliographystyle{JHEP}
\end{document}

%% file: macros.tex
\usepackage{ifthen}
\usepackage{tikz}
\usepackage{xspace}

\newcommand{\dNa}{\delta \eta_{N_\alpha}}
\newcommand{\etaNeq}{\eta_{N_\alpha}^{\rm eq}}
\newcommand{\etaNa}{\eta_{N_\alpha}}
\newcommand{\etaL}{\eta_L}
\newcommand{\za}{z_\alpha}
\newcommand{\mNa}{m_{N_\alpha}}
\newcommand{\GNa}{\Gamma_{N_\alpha}}
\newcommand{\HN}{H(z=1)}
\newcommand{\Tsph}{T_{\rm sph}}
\newcommand{\zsph}{z_{\rm sph}}

\newcommand{\heff}{h_{\rm eff}}
\newcommand{\geff}{g_{\rm eff}}
\newcommand{\deltah}{\delta_{h}}
\newcommand{\nn}{\nonumber}

\newcommand{\ie}{{i.e.}\xspace}
\newcommand{\eg}{{e.g.}\xspace}
\newcommand{\GeV}{{\rm GeV}\xspace}
\newcommand{\TeV}{{\rm TeV}\xspace}

\newcommand{\rhs}{RHS\xspace}
\newcommand{\lhs}{LHS\xspace}

\newcommand{\lrb}[1]{\left( #1 \right)}
\newcommand{\lrsb}[1]{\left[ #1 \right]}
\newcommand{\lrBigb}[1]{\Big( #1 \Big)}

\newcommand{\lrBiggcb}[1]{\Bigg\{ #1 \Bigg\}}

\newcounter{NumArgs}

\newcommand{\eqs}[1]{\setcounter{NumArgs}{0}\foreach\i in{#1}{\stepcounter{NumArgs}}%
	\ifthenelse{\equal{\theNumArgs}{1}}{(\ref{#1})}%
	{\ifthenelse{\equal{\theNumArgs}{2}}%
		{\foreach\i[count=\q]in{#1}{\ifthenelse{\equal{\q}{\theNumArgs}}{and (\ref{\i})}{(\ref{\i})~}}}%
		{\foreach\i[count=\q]in{#1}{\ifthenelse{\equal{\q}{\theNumArgs}}{and (\ref{\i})}{(\ref{\i}),~}}}}}

\newcommand{\Eqs}[1]{\setcounter{NumArgs}{0}\foreach\i in{#1}{\stepcounter{NumArgs}}%
	\ifthenelse{\equal{\theNumArgs}{1}}{Eq.~(\ref{#1})}%
	{\ifthenelse{\equal{\theNumArgs}{2}}%
		{Eqs.~\foreach\i[count=\q]in{#1}{\ifthenelse{\equal{\q}{\theNumArgs}}{and (\ref{\i})}{(\ref{\i})~}}}%
		{Eqs.~\foreach\i[count=\q]in{#1}{\ifthenelse{\equal{\q}{\theNumArgs}}{and (\ref{\i})}{(\ref{\i}),~}}}}}

\newcommand{\refs}[1]{\setcounter{NumArgs}{0}\foreach\i in{#1}{\stepcounter{NumArgs}}%
	\ifthenelse{\equal{\theNumArgs}{1}}{(\ref{#1})}%
	{\ifthenelse{\equal{\theNumArgs}{2}}%
		{\foreach\i[count=\q]in{#1}{\ifthenelse{\equal{\q}{\theNumArgs}}{and (\ref{\i})}{(\ref{\i})~}}}%
		{\foreach\i[count=\q]in{#1}{\ifthenelse{\equal{\q}{\theNumArgs}}{and (\ref{\i})}{(\ref{\i}),~}}}}}

\newcommand{\Figs}[1]{\setcounter{NumArgs}{0}\foreach\i in{#1}{\stepcounter{NumArgs}}%
	\ifthenelse{\equal{\theNumArgs}{1}}{Figure~\ref{#1}}%
	{\ifthenelse{\equal{\theNumArgs}{2}}%
		{Figures~\foreach\i[count=\q]in{#1}{\ifthenelse{\equal{\q}{\theNumArgs}}{and \ref{\i}}{\ref{\i}~}}}%
		{Figures~\foreach\i[count=\q]in{#1}{\ifthenelse{\equal{\q}{\theNumArgs}}{and \ref{\i}}{\ref{\i},~}}}}}

\newcommand{\Gen}[2]{\setcounter{NumArgs}{0}\foreach\i in{#2}{\stepcounter{NumArgs}}%
	\ifthenelse{\equal{\theNumArgs}{1}}{#1.~(\ref{#2})}%
	{\ifthenelse{\equal{\theNumArgs}{2}}%
		{#1.~\foreach\i[count=\q]in{#2}{\ifthenelse{\equal{\q}{\theNumArgs}}{and (\ref{\i})}{(\ref{\i})~}}}%
		{#1.~\foreach\i[count=\q]in{#2}{\ifthenelse{\equal{\q}{\theNumArgs}}{and (\ref{\i})}{(\ref{\i}),~}}}}}
